\documentclass{aa}  

\usepackage{graphicx}
\usepackage{txfonts}
\usepackage{lipsum}
\usepackage{subcaption}         
\usepackage{lscape}             
\usepackage{placeins}           

\begin{document}

   \title{Investigating X-ray Emission from Radio-detected Broad- and Narrow-line Seyfert 1 Galaxies}

   \author{Minhua Zhou\inst{1}\corrauth{zhoumh8@163.com}        
        \and Minfeng Gu\inst{2}\corrauth{gumf@shao.ac.cn}
        \and Lei Xu\inst{1}\email{lxu10.26@163.com}
        \and Mai Liao\inst{3}\email{mai.liao@mail.udp.cl}
        \and ZhengXue Chang\inst{4}\email{changzx520@163.com}
        }

   \institute{Jiangxi Province Key Laboratory of Applied Optical Technology, School of Mathematics and Physics, Shangrao Normal College, 401 Zhimin Road, Shangrao 334001, People's Republic of China
   \and Shanghai Astronomical Observatory, Chinese Academy of Sciences, 80 Nandan Road, Shanghai 200030, People’s Republic of China
   \and Instituto de Estudios Astrofisicos Facultad de Ingenieria y Ciencias Universidad Diego Portales Av. Eje`rcito 441, Santiago, Chile
   \and College of Mathematics and Physics, Handan University, Handan 056005, People’s Republic of China}

   \date{Received July 2, 20xx}
 
  \abstract 
   {This paper presents a study of the X-ray emission for a sample of radio-detected (RD-) broad- and narrow-line Seyfert 1 galaxies (NLS1s).}
   {To investigate the X-ray emission properties of early-stage accretion activity or early jets in NLS1s.} 
   {By fitting XMM-Newton X-ray spectra and combining with multi-band data from SDSS and FIRST, we conducted a statistical comparative study on the X-ray emission and multi-band correlations of NLS1s and broad-line Seyfert 1 galaxies (BLS1s).}
   {NLS1s and BLS1s show no significant difference in their soft X-ray blackbody temperatures, but NLS1s exhibit steeper hard X-ray photon indices and optical-to-X-ray slopes (\( \alpha_{\rm ox} \)). Moreover, correlation analyses of various parameters reveal intrinsic differences in the X-ray emission between the two samples. 
For instance, at comparable optical luminosities, NLS1s exhibit a higher \( \alpha_{\rm ox} \) (i.e., weaker X‑ray emission), marginally distinct dependencies between radio luminosity and black hole mass in the fundamental plane, and different correlations between \( \alpha_{\rm ox} \) and radio loudness. 
Based on the relationship between UV luminosity and \( \alpha_{\rm ox} \) in radio‑quiet (RQ-) active galactic nuclei (AGN), the estimated X‑ray excess (\( \Delta \alpha_{\rm ox} \)) in both samples is positively correlated with radio loudness. RQ-BLS1s generally display the X‑ray excess (\( \Delta \alpha_{\rm ox} < 0 \)), whereas RQ-NLS1s tend to show an X‑ray deficiency (\( \Delta \alpha_{\rm ox} > 0 \)). The Eddington ratio of RQ‑NLS1s derived from X‑ray luminosity is lower than that inferred from optical luminosity, further supporting their intrinsically weaker X‑ray emission.} 
   {These differences in X‑ray emission between NLS1s and BLS1s may ultimately stem from the earlier evolutionary stage of NLS1s. Multivariate linear fits among radio luminosity, X‑ray luminosity, and radio loudness for radio‑detected NLS1s and BLS1s support a multi‑wavelength radiative connection associated with jets.}


   \maketitle
   \nolinenumbers

\section{Introduction}

The canonical classification of active galactic nuclei (AGN) into Type 1 and Type 2 rests upon the detection of broad emission lines in their optical/UV spectra \cite[][]{1974ApJ...192..581K}. 
Type 1 AGN exhibit both broad and narrow lines, whereas only narrow lines are visible in Type 2 spectra. 
On the other hand, based on radio loudness—defined as the ratio of radio to optical luminosity, \cite[$R=f_{5\rm GHz}/f_{4400\rm \AA}$,][]{1989AJ.....98.1195K}—AGN are divided into two major categories: radio-loud ($R\ge 10$) and radio-quiet ($R<10$). These observational dichotomies are elegantly explained by the unified model \cite[][]{1993ARA&A..31..473A, 1995PASP..107..803U}, which posits an inclination-dependent obscuration by a circumnuclear dusty torus and the presence of a powerful relativistic jet.

The proximity of Seyfert galaxies offers a unique opportunity to probe the inner physics of AGN \cite[][]{2008ARA&A..46..475H}. 
Type 1 Seyfert galaxies are commonly subdivided into Narrow-Line Seyfert 1 (NLS1) and Broad-Line Seyfert 1 (BLS1) galaxies based on the width of their Balmer emission lines. 
The standard operational definition for NLS1s, established by \cite{1985ApJ...297..166O, 1989ApJ...342..224G}, includes: (1) a full width at half maximum (FWHM) of the H$\rm \beta$ broad component less than 2000 $\rm km\ s^{-1}$; (2) a flux ratio $[{\rm O\ III}] \lambda 5007 / H \beta < 3$; and (3) the presence of strong optical Fe II emission lines. 
It is noteworthy that the optical spectral properties of BLS1s and NLS1s do not exhibit a sharp dichotomy, suggesting a continuum of properties \cite[e.g.,][]{2006A&A...455..773V}.
In the X-ray band, NLS1s display distinctive characteristics compared to BLS1s, including a prominent and variable soft X-ray excess and steeper 2–10 keV spectra \cite[e.g.,][]{2011MNRAS.417.2426P}. The complex X-ray spectra of NLS1s may originate from several mechanisms, such as relativistic reflection from the accretion disk \cite[e.g.,][]{2004MNRAS.353.1071F}, partial-covering absorption by neutral or ionized material \cite[e.g.,][]{2004MNRAS.349L...7G, 2004PASJ...56L...9T}. 
The systematic differences between BLS1s and NLS1s above cannot be explained solely by orientation, but are more likely to reflect fundamental differences in black hole mass, accretion rate, corona, and jet formation. 
The defining properties of NLS1 galaxies—their low black hole masses and high accretion rates \cite[e.g.,][]{2002ApJ...565...78B, 2004A&A...426..797C}—imply that they may be young AGN in an early, actively growing stage of evolution \cite[e.g.,][]{2008RMxAC..32...86K, 2011MNRAS.417.2426P, 2012AJ....143...83X}. 

Notably, although most NLS1s are radio-quiet, a radio-loud population exists \cite[$\sim 7 \%$][]{2006AJ....132..531K}. Some of these radio-loud NLS1s (RL-NLS1s) possess powerful relativistic jets \cite[e.g.,][]{2008ApJ...685..801Y, 2015ApJS..221....3G, 2015A&A...575A..13F}, an assertion further strengthened by multiple detections of gamma-ray emission from such sources \cite[][]{2009ApJ...707L.142A, 2018ApJ...853L...2P}. 
As shown in \cite{2006AJ....132..531K}, most RL-NLS1s in their sample are compact, steep-spectrum sources, and hence are likely associated with compact steep-spectrum (CSS) radio sources, sharing the same radio properties as young radio AGNs \cite[][]{2010AJ....139.2612G, 2014MNRAS.441..172C, 2015ApJS..221....3G}.

In our previous work, we studied the X-ray difference between radio-loud quasars (RLQs) and radio-quiet quasars (RQQs) by building the composite X-ray spectra for the Sloan Digital Sky Survey (SDSS) quasars and 3CRR quasars \cite[][]{2020ApJ...893...39Z, 2021RAA....21....4Z}. 
After excluding blazars, we found that the radio-loud X-ray composite spectra of 3CRR quasars and SDSS quasars are similar to that of the RLQ sample in \citet{2011ApJS..196....2S}, but differ significantly from that of the RQQ sample.
Moreover, by dividing the radio-detected quasars into radio-loudness bins, \citet{2021RAA....21....4Z} found a general increase of the X-ray composite spectra with increasing radio loudness, a significant correlation between the optical-to-X-ray spectral index and radio loudness, and a unified multi-correlation among radio luminosity, X-ray luminosity, and radio loudness.
These results could be possibly explained by the corona-jet model, in which the corona and jet are directly related. 

\cite{2018rnls.confE..15K} summarized previous studies about the multi-wavelength observational properties of RL-NLS1s, suggesting that RL-NLS1s host relatively low-mass black holes with near-Eddington accretion rates. These systems are capable of producing relativistic jets, and their observed emission is jointly contributed by the accretion corona and the jet. Given the young nature of NLS1s—whether in terms of accretion or jet activity—such a comparative analysis is well motivated. It can not only reveal the jet–corona interaction under high accretion rates, but also test key frontier issues, including the AGN unification model, the disk–jet connection, and the origin of soft X-ray excess. 
To further investigate the impact of jet activity on X-ray emission, this paper presents a comparative analysis of the X-ray properties of radio-detected NLS1s (RD-NLS1s) and BLS1s. Historically, extensive comparative studies of X-ray emission between NLS1s and BLS1s have been conducted \cite[][]{1996A&A...305...53B, 2010ApJ...713L..11Z, 2010ApJS..187...64G, 2011ApJ...727...31A, 2011MNRAS.417.2426P, 2018MNRAS.478.4068L, 2020MNRAS.491..532G, 2020MNRAS.498.5207W, 2025MNRAS.539..832J}; however, most have focused on radio-quiet NLS1s (RQ-NLS1s), while RD-NLS1s have been largely overlooked. 
Most investigations of RD-NLS1s have largely focused on multi-epoch, broadband spectral analyses of individual sources \cite[e.g.,][]{2015A&A...574A.121K, 2018MNRAS.479.2464G, 2020MNRAS.496.2922M, 2021MNRAS.507.3572X, 2025A&A...698A.160R, 2026MNRAS.548ag588Z}, or on extreme radio-loud (gamma-ray emitting) NLS1 samples \cite[e.g.,][]{2019ApJ...872..169P, 2019A&A...632A.120B, 2020MNRAS.496.2213D, 2022RAA....22i5006C, 2023MNRAS.523..441Y, 2026ApJ...996..118C}. Nevertheless, the overall sample size of X-ray studies of RD-NLS1s remains relatively small \cite[e.g.,][]{2006AJ....132..531K}. 

\cite{2024MNRAS.527.7055P} constructed a new NLS1 catalog based on the Sloan Digital Sky Survey Data Release 17 (SDSS-DR17) spectral data, containing 22,656 NLS1 galaxies—twice the number in the previous SDSS-DR12 sample \cite[][]{2017ApJS..229...39R}—thereby providing an ideal sample for studying the X-ray emission properties of RD-NLS1s. 
Our sample is presented in Section \ref{sec:sample}, and the optical and X-ray data are shown in Section \ref{sec:data}. 
In Section \ref{sec:results}, we provide the main results of X-ray spectra for our radio-detected NLS1s and BLS1s. 
The results are discussed in Section \ref{sec:discc} and summarized in Section \ref{sec:sum}. 

In this work, we adopt the cosmology parameters $H_{0}=70\rm ~km~ s^{-1}~Mpc^{-1}$, $\Omega_{\rm m}=0.3$ and $\Omega_{\Lambda}=0.7$. 
Photon index $\Gamma$ of power-law is defined by 
$A(E) = K E^{-\Gamma}$, where $K$ is photons at $1~\rm keV$ and $E$ is photon energy.
The spectral index $\alpha$ is defined as $f_{\rm \nu} \varpropto \nu^{-\alpha}$ with $f_{\rm \nu}$ being the flux density at frequency $\rm \nu$. 

\section{Sample selection} \label{sec:sample}

To construct a large RD-NLS1 sample and a comparison BLS1 sample, we started from the current largest optical-spectra-selected NLS1 and BLS1 samples based on the SDSS-DR17 data \cite[][]{2024MNRAS.527.7055P}, which include 22,656 NLS1s and 52,273 BLS1s. To obtain the radio-detected galaxies, we cross-matched the primary NLS1s and BLS1s with the Faint Images of the Radio Sky at Twenty-Centimeters (FIRST) Survey Catalog \citep{1997ApJ...475..479W, 2015ApJ...801...26H} within a $5\arcsec $ radius.
The FIRST-detected galaxies were then cross-matched with the 4XMM-DR14 spectral-fit database \citep[XMMFITCAT,][]{2020A&A...641A.136W} using a standard $5\arcsec $ matching radius.
We prefer to use the data from the X-ray Multi-Mirror Mission (XMM-Newton) because it has a large collecting area and typically provides long, uninterrupted exposures, resulting in highly sensitive observations.
Finally, blazars in the BZCAT catalog \citep{2009A&A...495..691M,2015Ap&SS.357...75M} Edition 5.0.0 were checked and then excluded to minimize the significant jet contribution in the X-ray band due to strong beaming effects. These steps yield a sample of radio-detected galaxies consisting of 60 NLS1s and 163 BLS1s. 
Given the non-simultaneity of the multi-band observations, we exclude sources exhibiting significant long-term variability, for instance, the changing-look or changing-state AGNs SDSS J015105.79-003426.4 \cite[][]{2025ApJ...994..154L}, SDSS J020514.77-045639.7 \cite[][]{2022ApJ...933..180G}, SDSS J020615.99-001729.1 \cite[Mrk 1018, e.g.,][]{2016A&A...593L...9H, 2021MNRAS.506.4188L}, SDSS J022014.57-072859.3 \cite[][]{2020MNRAS.491.4925G}, SDSS J082942.66+415436.8 \cite[][]{2021A&A...650A..33P}, SDSS J141759.54+250812.7 \cite[NGC 5548, e.g.,][]{2019ApJ...882L..30D} and SDSS J161413.20+260416.2 \cite[][]{2020MNRAS.491.4925G}.
With the X-ray process in Section \ref{subsec:x-ray}, some objects are located outside the CCD chip or heavily affected by background or weak signals (photon counts less than 250), leaving only 44 NLS1s and 125 BLS1s. Our complete sample is shown in Table \ref{tableS1.tab}. 

Following the definition of radio loudness $R=f_{\rm 5GHz} /f_{\rm 4400\AA}$, we converted FIRST $1.4~\rm GHz$ flux to rest-frame $5~\rm GHz$ flux by assuming a radio spectral index of $\alpha=0.5$, and estimated the rest-frame $\rm 4400~\AA$ flux from the power-law component derived by fitting SDSS spectra (see Section \ref{subsec:optical}). 
The rest-frame luminosities (at $4400~\rm \AA$ and $2500~\rm \AA$) and the radio loudness are listed in the last three columns of Table \ref{tableS1.tab}. 

\begin{table*}[ht!]
\caption{\label{tableS1.tab}Radio-detected NLS1 and BLS1 samples}
\centering
\begin{tabular}{c|ccccccccr}
\hline\hline
\multicolumn{1}{c}{Name} & \multicolumn{1}{c}{R.A.} & \multicolumn{1}{c}{Dec.} & \multicolumn{1}{c}{$z$} & \multicolumn{1}{c}{Plate} & \multicolumn{1}{c}{MJD} & \multicolumn{1}{c}{Fiber} & \multicolumn{1}{c}{$\nu L_{\nu,4400}$} & \multicolumn{1}{c}{$\nu L_{\nu,2500}$} & \multicolumn{1}{c}{$R$} \\
\multicolumn{1}{c}{SDSS J} & \multicolumn{1}{c}{$\rm deg$} & \multicolumn{1}{c}{$\rm deg$} & \multicolumn{1}{c}{ } & \multicolumn{1}{c}{ } & \multicolumn{1}{c}{ } & \multicolumn{1}{c}{ } & \multicolumn{1}{c}{$\rm erg~s^{-1}$} & \multicolumn{1}{c}{$\rm erg~s^{-1}$} & \multicolumn{1}{c}{ } \\
\multicolumn{1}{c}{(1)} & \multicolumn{1}{c}{(2)} & \multicolumn{1}{c}{(3)} & \multicolumn{1}{c}{(4)} & \multicolumn{1}{c}{(5)} & \multicolumn{1}{c}{(6)} & \multicolumn{1}{c}{(7)} & \multicolumn{1}{c}{(8)} & \multicolumn{1}{c}{(9)} & \multicolumn{1}{c}{(10)} \\
\hline
\multicolumn{10}{c}{NLS1 sample} \\
\hline
011448.67-002946.0 & 18.703  & -0.496  & 0.034  &  0398 &  51789 &  0261 & 42.43  & 42.83  & 33.06 \\ 
013955.76+061922.5 & 24.982  & 6.323  & 0.395  &  11071 &  58429 &  0123 & 45.30  & 45.42  & 0.69 \\ 
030313.02-001457.5 & 45.804  & -0.249  & 0.700  &  0802 &  52289 &  0148 & 44.99  & 45.97  & 4714.41 \\ 
\multicolumn{10}{c}{...} \\ 
\hline\noalign{\smallskip}
\multicolumn{10}{c}{BLS1 sample} \\
\hline
000710.01+005329.0 & 1.792  & 0.891  & 0.316  & 9407 & 58041 & 0415 & 44.90  & 44.55  & 1.93 \\ 
001137.24+144201.4 & 2.905  & 14.700  & 0.132  & 0752 & 52251 & 0231 & 43.17  & 44.15  & 10.88 \\ 
004607.97+090720.9 & 11.533  & 9.122  & 0.238  & 11047 & 58424 & 0120 & 45.01  & 44.94  & 0.62 \\ 
\multicolumn{10}{c}{...} \\ 
\hline
\end{tabular}
\tablefoot{In this Table, Column (1): SDSS name; Column (2): Right ascension in degree (J2000); Column (3): Declination in degree (J2000); Column (4): Redshift; Column ($5-7$): Plate, MJD, and Fiber for SDSS spectra; Column ($8-9$): The luminosity at rest frame $4400\rm \, \AA$ and $2500\rm \, \AA$; Column (10): Radio loudness. 
Table \ref{tableS1.tab} is published in its entirety in the machine-readable format.}
\end{table*}

\section{Data reduction} \label{sec:data}

The XMM-Newton satellite is equipped with both optical/UV and X-ray detectors, enabling excellent multi-band quasi-simultaneous observations. For optical data, this study focuses solely on SDSS spectroscopy, as optical/UV photometry observed by the XMM-Newton telescope of Seyfert galaxies is likely contaminated by host galaxy emission, making clean separation difficult. In contrast, SDSS spectral fitting can isolate the nuclear AGN emission with relatively high accuracy.

Some sources were observed multiple times by the SDSS survey; we selected the same observational data as listed in the table of \cite{2024MNRAS.527.7055P}. For sources with multiple XMM-Newton observations, we prioritized those closest in time to the SDSS observations. If all X-ray observations significantly differed from the SDSS epoch (by more than one year), we instead selected higher-quality observations—those with longer exposure times, fewer pile-up effects, or lower background contamination.
For some targets observed repeatedly but with short individual exposures, we combined these observations for joint fitting (e.g., the BLS1 SDSS J113702.86+154656.8).

\subsection{Optical} \label{subsec:optical}

\cite{2024MNRAS.527.7055P} provided decomposition fitting results for SDSS spectra, including emission-line information such as {H\,{\footnotesize $ \alpha $}}, {H\,{\footnotesize $\beta$}}, and {[O\,{\footnotesize III}]}, as well as detailed parameters such as black hole mass and Eddington luminosity, but lacked information on the nuclear continuum spectral components. To estimate radio loudness and the optical/UV–X-ray slope, we employed the latest Python-based program PyQSOfit \cite[][]{2018ascl.soft09008G, 2024ApJ...974..153R} to perform spectral component analysis on each SDSS spectrum, separating the host galaxy contribution and extracting the nuclear continuum flux. Our specific fitting procedure follows \cite{2021RAA....21....4Z}, with key fitting steps outlined below.

The SDSS optical spectra were firstly corrected for Galactic extinction with the reddening map of \cite{1998ApJ...500..525S}.
A single power-law, host galaxy component \cite[][]{2004AJ....128..585Y, 2004AJ....128.2603Y} and {Fe\,{\footnotesize II}} emission \cite[][]{1992ApJS...80..109B, 2001ApJS..134....1V} were applied to fit the continuum at several line-free windows \cite[see e.g.,][]{2011ApJS..194...45S}. The flux at $\rm 4400\,\AA$ and $\rm 2500\,\AA$ at source rest frame was then estimated from the fitted power-law continuum. 

\subsection{X-ray} \label{subsec:x-ray}

For all our sample sources, the XMMFITCAT provides spectral fitting results using various simple models (absorbed power-law, absorbed thermal, and absorbed blackbody) and complex models (absorbed thermal plus power-law, absorbed double power-law, and absorbed blackbody plus power-law). However, we decided to reduce the XMM-Newton data for all galaxies on our own, rather than directly adopting the results from XMMFITCAT, in order to ensure a uniform spectral analysis.

We processed the data with XMM-Newton Scientific Analysis Software (SAS) package step by step with SAS cookbook same as \cite{2021RAA....21....4Z}. The pn and MOS datasets were processed with \texttt{epproc} and \texttt{emproc} in SAS-21.0.0, and were filtered in energy ranges of $0.2-15.0~\rm keV$ and $0.2-12.0~\rm keV$, respectively. We created and checked the light curve to filter out the time interval with large flares. 
The source X-ray spectrum was then extracted from the pn data from a source-centered radius of $32\arcsec$ with a background source-free region of $40\arcsec$ around the object. If the source locates at the edge of the pn CCD chip or is affected by other source, we accordingly decreased the radius of source region (e.g., $24\arcsec$, or even $20\arcsec$). In the case that the source exactly locates at the gap or is out of pn chips, we extracted the X-ray spectrum from MOS data. 
Pile-up effect was checked with \texttt{epatplot}. 
To mitigate pile-up in affected targets, source spectra are extracted from an annular region with an inner radius of $7.5\arcsec$ and an outer radius of $40\arcsec$. 
Finally, all spectra were rebinned with a minimum of 15 counts for background-subtracted spectral channel and oversampling the intrinsic energy resolution by a factor of no larger than three. 

All X-ray spectra were fitted using XSPEC v12.14 with the $chi-squared$ statistical method. To search for soft X-ray excess, we first fitted the X-ray spectrum in the $2.0-12.0~\rm keV$ band with a Galactic-absorbed single power-law model (\textit{phabs $\times$ powerlaw}; \citealt{2016A&A...594A.116H}), and then visually inspected the residuals between the model and the data in the soft X-ray band $0.3-2.0~\rm keV$. If the residuals showed a prominent excess, we re-fitted the spectrum with an absorbed blackbody plus power-law model (\texttt{phabs $\times$ zphabs $\times$ (bbody $+$ powerlaw)}). Otherwise, we compare the results of fitting the $0.3-12.0~\rm keV$ spectra using either a single power-law model (\texttt{phabs $\times$ zphabs $\times$ powerlaw}) or a two-component model (blackbody $+$ power-law), and apply the F-test to evaluate the necessity of adding the soft X-ray excess component. 

Most X-ray spectra resemble quasar spectra \cite[][]{2021RAA....21....4Z} and can be successfully fitted with a single absorbed power-law or a blackbody plus power-law model. However, some spectra in this work exhibit additional absorption features below $1\ \rm keV$, which may originate from partial covering absorption (\texttt{zxipcf}) or an extra absorption edge at $0.6-0.7\ \rm keV$ \cite[\texttt{zedge}, e.g.,][]{1999ApJS..125..317L, 2006MNRAS.365.1067C}. 
Intrinsic partial-covering absorption may readily mimic or mask a soft excess component. To assess this, we compared a single zxipcf-absorbed power-law model against a two-component model (blackbody + power-law) using the Akaike Information Criterion (AIC) to identify the preferred fitting model. 
The observed $2-10~\rm keV$ fluxes of primary power-law components were estimated by \texttt{cflux}. 
The main components of our best-fit models are shown in Table \ref{tableS2.tab} in Appendix \ref{appendix1}. 

\begin{figure}[htb!]
	\centering
	\includegraphics[width=0.4\textwidth]{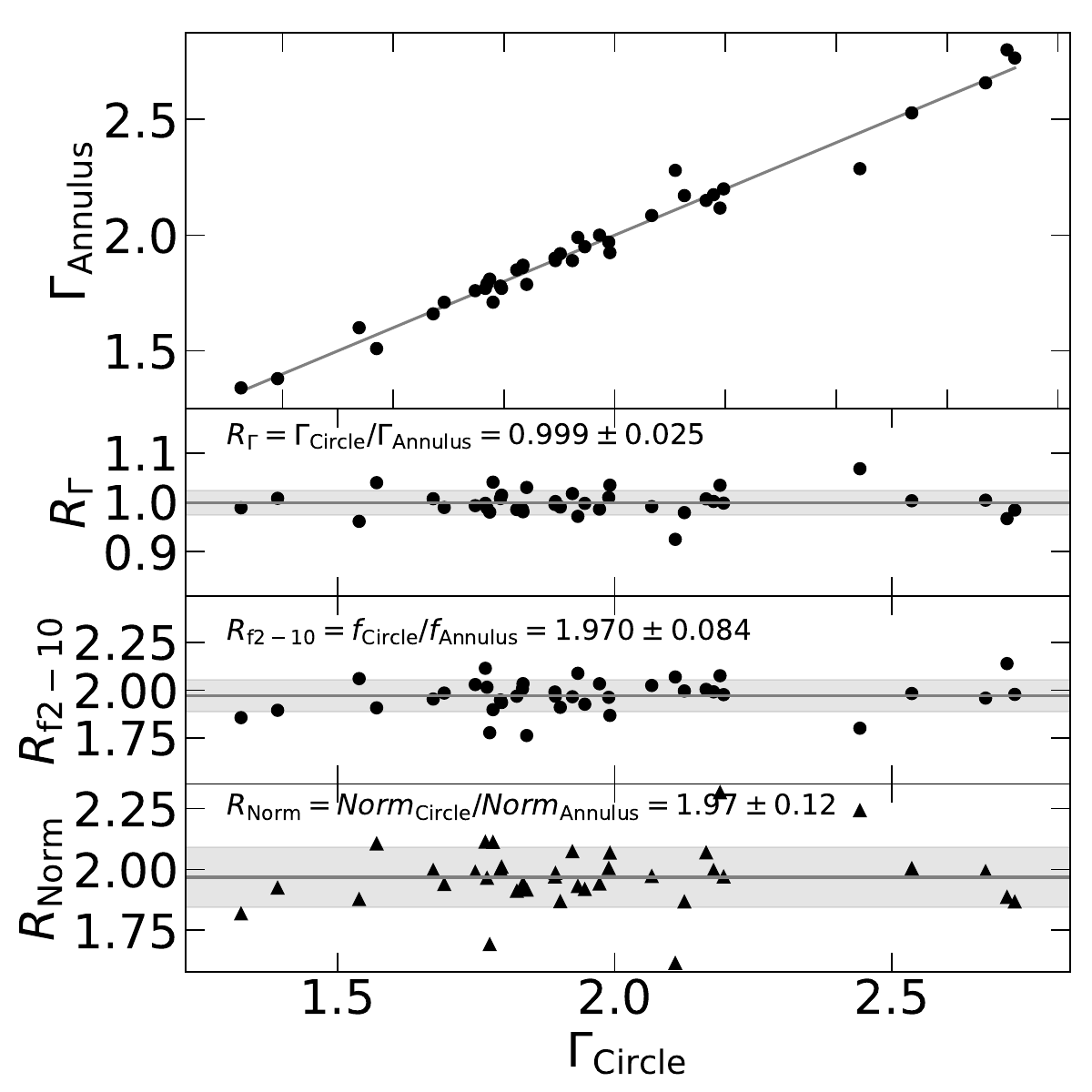}
	\caption{Comparison of power-law spectral parameters across different regions. The top panel compares the photon indices from the circular and annular regions. The black solid line represents the locus where the photon indices of the two regions are equal. The lower three panels show the ratios (circular/annular) of the photon index, $2-10\rm keV$ flux, and the power-law normalization, respectively. The black solid line and gray-shaded region correspond to the mean and the 1$\sigma$ uncertainty of the ratios in each panel.  \label{fig:gamma_an-cc}}
\end{figure}

The source spectra observed by XMM-Newton in different regions exhibit similar point spread functions for soft and hard X-rays, but show significant differences in flux values \cite[e.g.,][]{2023MNRAS.519..909Z}. Consequently, the fluxes from pile-up-affected sources extracted from annular regions differ from those obtained from circular regions. 
All pile-up-affected sources in our sample (8 NLS1s and 4 BLS1s) are concentrated near the center of the XMM-Newton pn detector. 
To investigate this effect, we selected several bright sources (10 NLS1s and 21 BLS1s, 38 observations) located near the center of the XMM-Newton pn detector field of view, which are unaffected by pile-up, and compared source spectra from both circular ($32\arcsec$) and annular ($7.5\arcsec-40\arcsec$) regions. 
From fitting the X-ray spectra of all observations, we find that the same model can be applied to both the circular and annular regions in each case. Of these, 10 observations are well described by a single absorbed power-law spectrum, while the remaining 28 exhibit a soft X-ray excess that requires an additional blackbody component. Fig. \ref{fig:gamma_an-cc} presents a comparison of the power-law parameters derived from the fits for different regions in each observation. Our results show good consistency between the two regions in terms of the photon index of the power-law component and the temperature of the blackbody component.
The other spectral fitting parameters for the annular region can be well accounted for by scaling constants, including the power-law normalization ratio $R_{\rm Pnorm} = Pnorm_{\rm circle} / Pnorm_{\rm annulus} = 1.97 \pm 0.12$, the blackbody normalization ratio $R_{\rm Bnorm} = Bnorm_{\rm circle} / Bnorm_{\rm annulus} = 1.86 \pm 0.34$, and the $2-10~\rm keV$ flux ratio $R_{\rm flux} = f_{\rm circle} / f_{\rm annulus} = 1.970 \pm 0.084$. These ratios were used to obtain accurate $2~\rm keV$ fluxes and $2-10~\rm keV$ fluxes for the pile-up affected spectra in the following analysis.

\section{Results} 
\label{sec:results}

\subsection{Spectral fitting results} 
\label{subsec:spec_results}

A statistical comparison of the X-ray spectral fitting results for the NLS1 and BLS1 samples (as shown in Table \ref{tableS2.tab} in Appendix \ref{appendix1}) shows that, of the 44 NLS1 X-ray spectra, 9 can be fitted with a single power-law model, 34 exhibit a soft X-ray excess, and 1 present ultra-soft X-ray spectra (fitted with a single blackbody model); by contrast, among the 125 BLS1 X-ray spectra, 61 can be fitted with a single power-law model, and 64 exhibit a soft X-ray excess component. 
Our RD-NLS1s exhibit a higher soft X-ray excess detection rate compared to BLS1s, with rates of 79.5\% and 51.2\%, respectively. 
An analysis based on radio loudness shows that, among 24 RL-NLS1s and 20 RQ-NLS1s, the numbers of sources with soft X-ray excess are 16 and 19, corresponding to detection rates of 66.7\% and 95.0\%. 
RQ-NLS1s exhibit a detection rate of nearly 100\% for the soft X-ray excess, consistent with the results of \cite{2020MNRAS.491..532G}. 
In contrast, among 74 RL-BLS1s and 51 RQ-BLS1s, the numbers of sources exhibiting soft X-ray excess are 34 and 30, with detection rates of 45.9\% and 58.8\%, respectively. When categorized by radio loudness, both RQ-NLS1s and RQ-BLS1s show higher soft X-ray excess detection rates than their radio-loud counterparts. Our results are consistent with previous findings \cite[][]{2021RAA....21....4Z}, which reported that the soft X-ray excess detection rate in RQQs (60.0\%) is higher than that in RLQs (25.0\%). 
However, among radio-detected sources, the soft X-ray excess detection rate of NLS1s is higher than that of BLS1s and quasars, indicating the possible distinct nature of the X-ray emission in NLS1s. 

\begin{figure*}[htb!]
    \sidecaption
	\includegraphics[width=6cm]{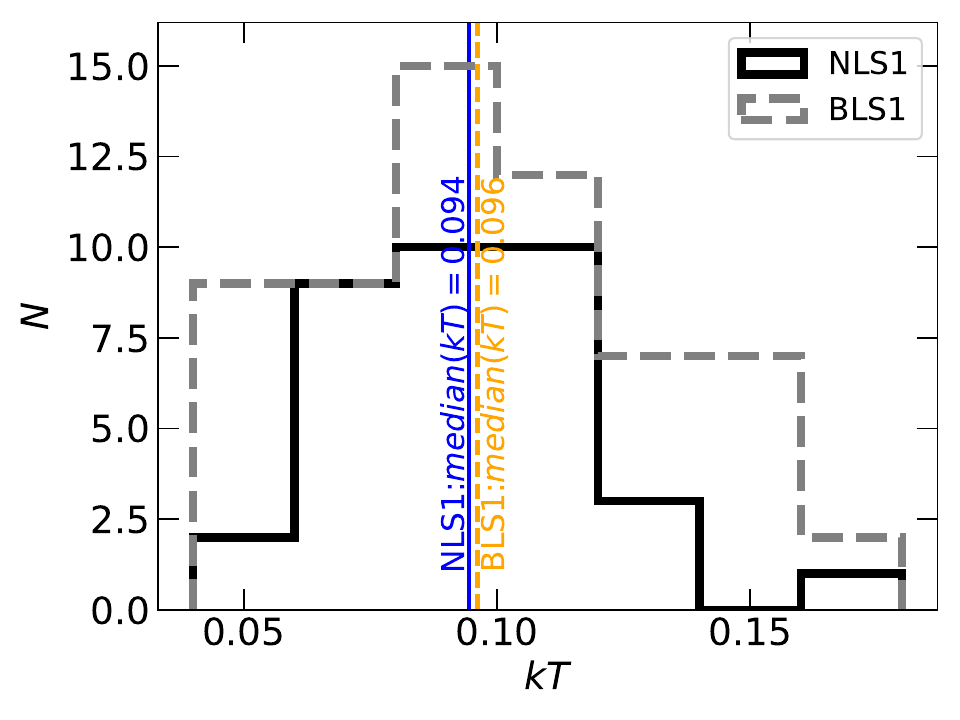}
    \includegraphics[width=6cm]{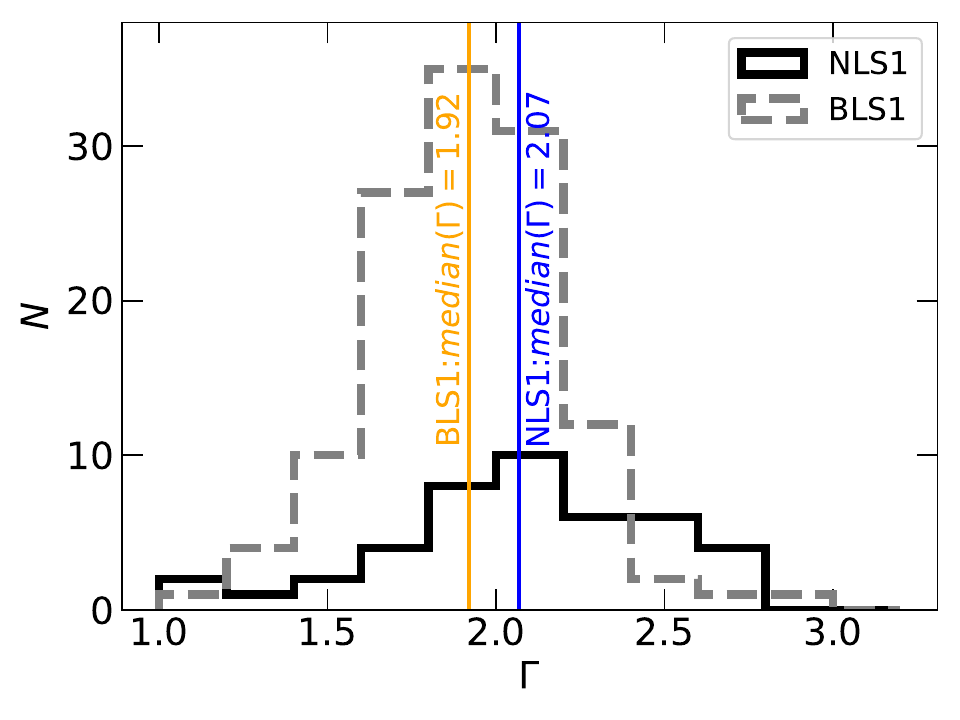}
	\caption{Distribution of blackbody temperature (left panel) and power-law photon index (right panel) for NLS1s and BLS1s. The blue and orange thick lines mark the median values of the correlated parameter for NLS1s and BLS1s, respectively.   \label{fig:kT_gamma}}
\end{figure*}

Fig. \ref{fig:kT_gamma} compares the distributions of the soft X-ray excess component (blackbody temperature) and the power-law component (photon index) between the two samples. The blackbody temperature distributions are similar, with a Kolmogorov–Smirnov (KS) test yielding $p = 0.29$. 
Considering the standard error ($\frac{\sigma}{\sqrt{N}}$), NLS1s exhibit comparable blackbody temperatures to those of BLS1s with median values of $0.094 \pm 0.004$ and $0.096 \pm 0.005$, respectively.
In contrast, the photon indices of the power-law component differ significantly between the two samples (KS test $p = 3.74 \times 10^{-3}$). The primary X-ray spectra of NLS1s are steeper than those of BLS1s, although the difference is less pronounced than reported in previous studies \cite[][]{2020MNRAS.491..532G, 2020MNRAS.498.5207W}. Considering the standard error, their median values are $2.07 \pm 0.06$ and $1.92 \pm 0.03$, respectively.

Regarding X-ray absorption, most of the NLS1s and BLS1s in our sample show no significant features. Excluding the upper limits from the neutral hydrogen column density fits, only 12 of the 44 NLS1s have $n_{\rm H} > 10^{20} \, \rm cm^{-2} $ (with 8 having $n_{\rm H} > 10^{21} \, \rm cm^{-2} $ and 2 showing partial-covering absorption features modeled by $zxipcf$), while only 19 of the 125 BLS1s exceed this threshold (with 13 having $n_{\rm H} > 10^{21} \, \rm cm^{-2} $ and 4 showing partial-covering absorption features via $zxipcf$). The fraction of NLS1s with high X-ray absorption (29.5\%) is slightly larger than that of BLS1s (15.2\%).

\subsection{$\alpha_{\rm ox}$ comparison}
\label{subsec:aox}

\begin{figure}[htb!]
	\centering
	\includegraphics[width=0.4\textwidth]{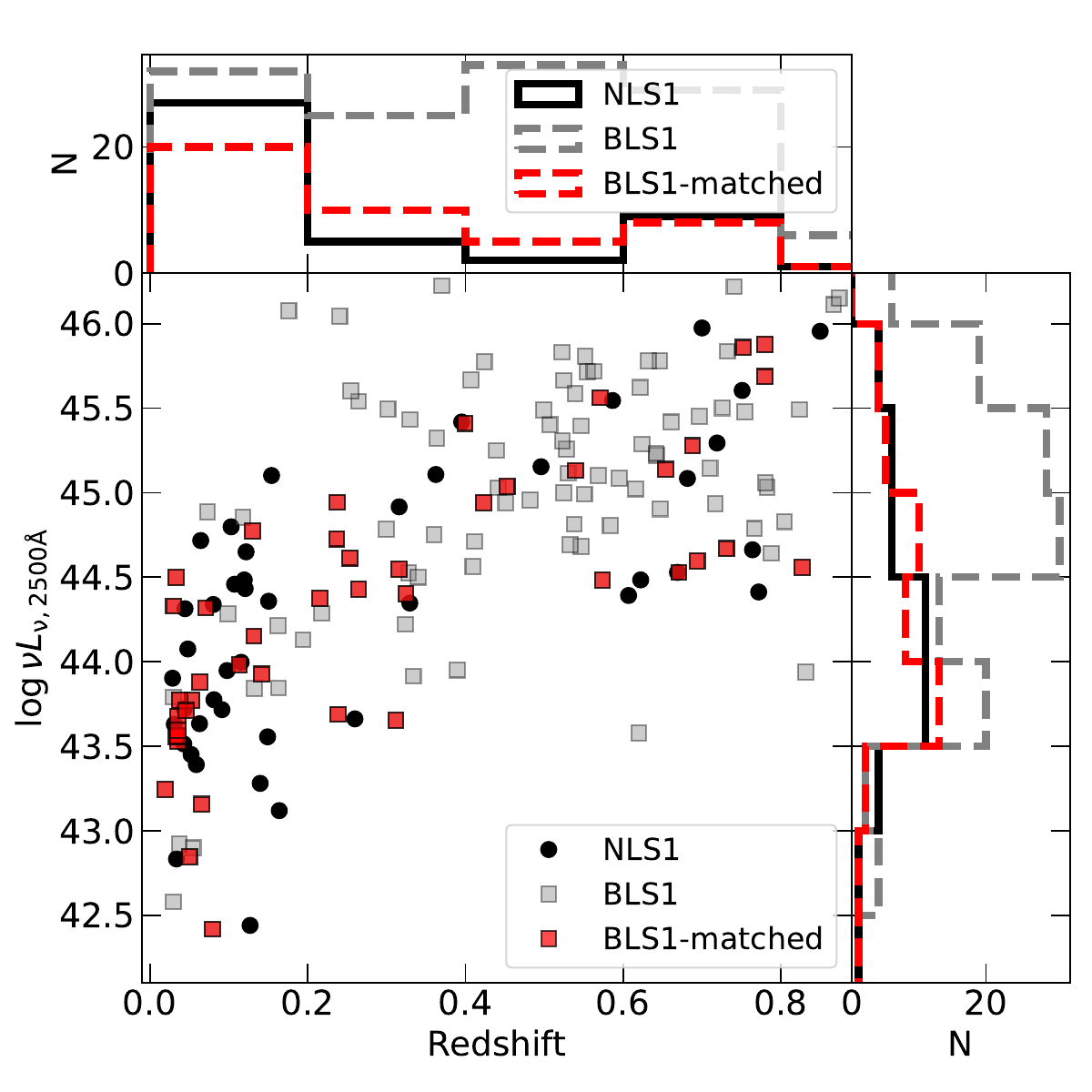}
	\caption{The distribution of redshift and $2500~\rm \AA$ luminosity for the NLS1 and matched BLS1 samples. \label{fig:NL-BL}}
\end{figure}

Due to the contamination from the optical/UV emission of the host galaxies in NLS1s, we are unable to construct an optically-normalized X-ray composite spectrum following the approach of \cite{2021RAA....21....4Z}. To enable a comparison between RD-NLS1s and BLS1s, we instead employ SDSS spectral fitting to subtract the host galaxy contribution, and then estimate the optical/UV-to-X-ray spectral slope $\alpha_{\rm ox}$ (defined as $\alpha_{\rm ox} = -\frac{\log(L_{\nu,2\rm keV}/L_{\nu,2500\rm \AA})}{\log(\nu_{2\rm keV}/\nu_{2500})}=-0.384\log(L_{\nu,2\rm keV}/L_{\nu,2500\rm \AA})$). A larger $\alpha_{\rm ox}$ value indicates a steeper spectral slope between $2~\rm keV$ and $2500~\rm \AA$ \cite[i.e., weaker X-ray emission relative to the optical/UV flux, see Fig. 1 in][]{2025ApJ...995..197C}.
Given that differences between the samples \cite[e.g., higher luminosity tends to yield statistically larger $\alpha_{\rm ox}$ values][and their references]{2015A&ARv..23....1B} may introduce additional selection effects on the results, we constructed a BLS1 subsample that is matched one-to-one with NLS1s in the redshift–luminosity plane, using the full BLS1 sample as the parent population. 
As shown in Fig. \ref{fig:NL-BL}, there are significant differences in the redshift and $2500~\rm \AA$ luminosity distributions between the NLS1 and BLS1 samples, with BLS1s exhibiting higher values in both quantities. In contrast, after matching, the matched BLS1s (Fig. \ref{fig:NL-BL}) has redshift and $2500~\rm \AA$ luminosity distributions indistinguishable from those of the NLS1s, with K‑S test $p$ values of $0.80$ and $0.99$, respectively.

\begin{figure*}[htb!]
	\sidecaption
	\includegraphics[width=12cm]{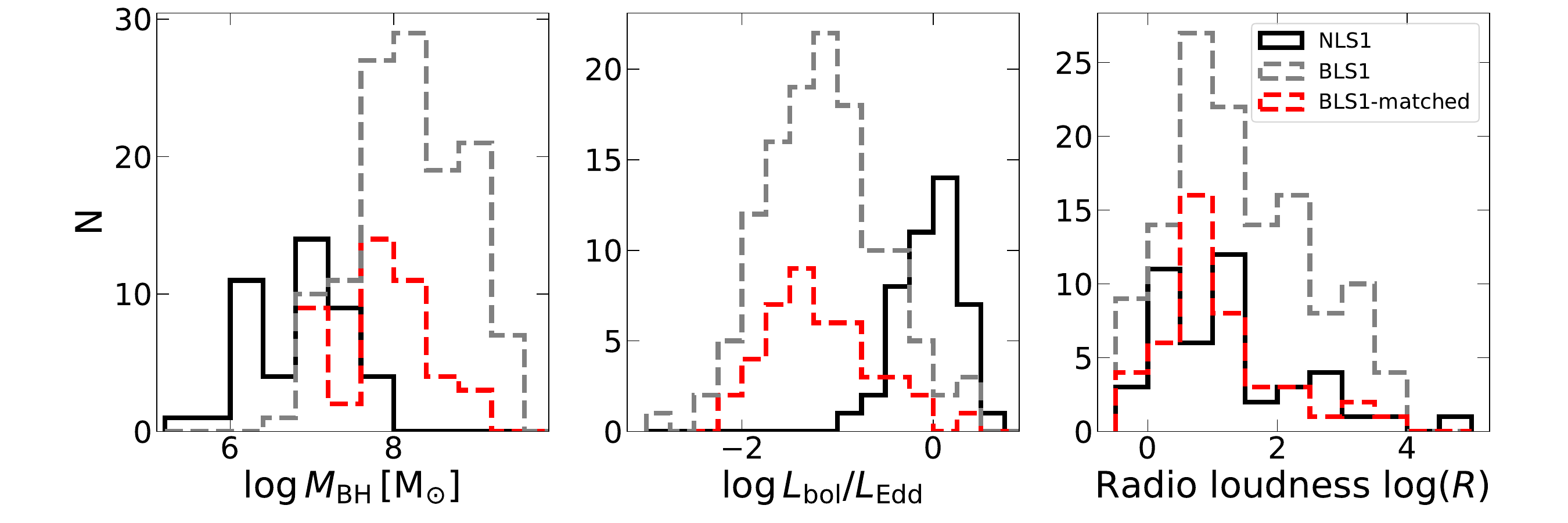}
	\caption{The distributions of black hole mass ($left$), Eddington ratio ($middle$), and radio loudness ($right$) for our NLS1, BLS1 and matched BLS1 samples.  \label{fig:hist-MER}}
\end{figure*}

Fig. \ref{fig:hist-MER} compares the distributions of black hole mass, Eddington ratio, and radio loudness across the three samples. Compared with NLS1s, the BLS1s and matched BLS1s samples exhibit statistically higher black hole masses and lower Eddington ratios (KS test $p$-values $<< 0.05$ for both), while their radio loudness distributions are similar (KS test $p = 0.22$ for BLS1s vs. NLS1s, and $p = 0.47$ for matched BLS1s vs. NLS1s).

\begin{figure}[htb!]
	\centering
	\includegraphics[width=0.4\textwidth]{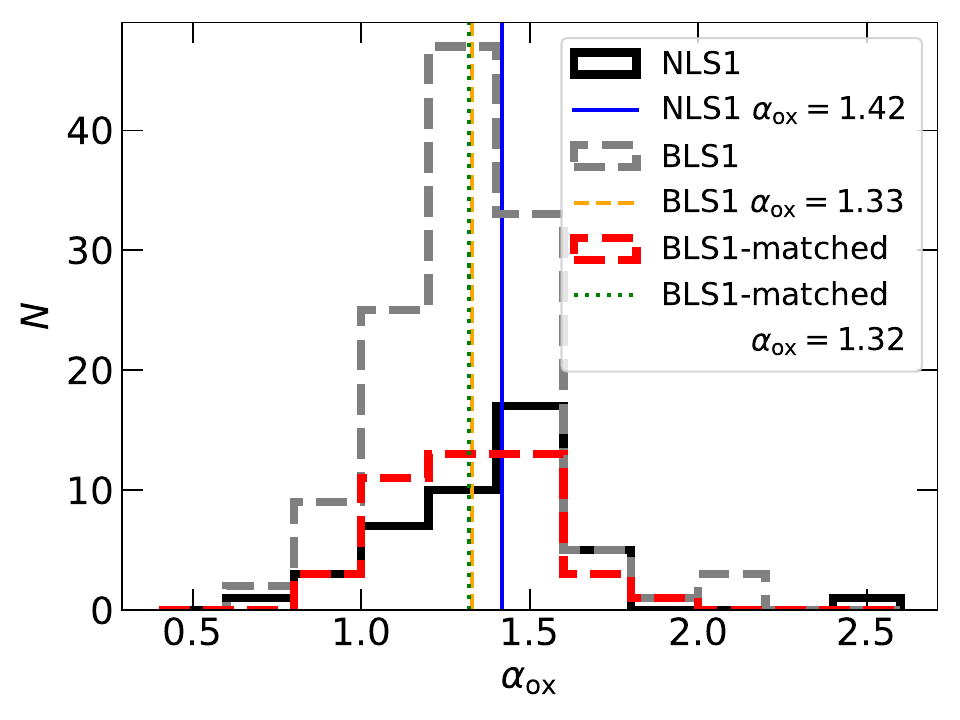}
	\caption{Distributions of $\alpha_{\rm ox}$ for our NLS1, BLS1, and matched BLS1 samples. The blue solid line, orange dashed line, and green dotted line indicate the median $\alpha_{\rm ox}$ of each sample, respectively. \label{fig:hist-aox}}
\end{figure}

NLS1s and BLS1s exhibit a slight difference in their optical-to-X-ray spectral slope $\alpha_{\rm ox}$ (KS test $p = 0.013$; Figure \ref{fig:hist-aox}). The median $\alpha_{\rm ox}$ values for the two samples are $1.42\pm0.04$ and $1.33\pm0.02$, respectively. After correcting for potential selection effects, the matched BLS1 and BLS1 samples show a similar median $\alpha_{\rm ox}$ of $1.32\pm0.03$, and their distributions are statistically consistent (KS test $p = 0.99$). 
Our median $\alpha_{\rm ox}$ values are in good agreement with those found by \cite{2010A&A...512A..34L, 2023MNRAS.520.2781K} with comparable optical luminosity distributions.

\subsection{$\alpha_{\rm ox}$ correlations}

\subsubsection{$\alpha_{\rm ox}$ versus optical/UV luminosity}

\begin{figure}[htb!]
	\centering
	\includegraphics[width=0.4\textwidth]{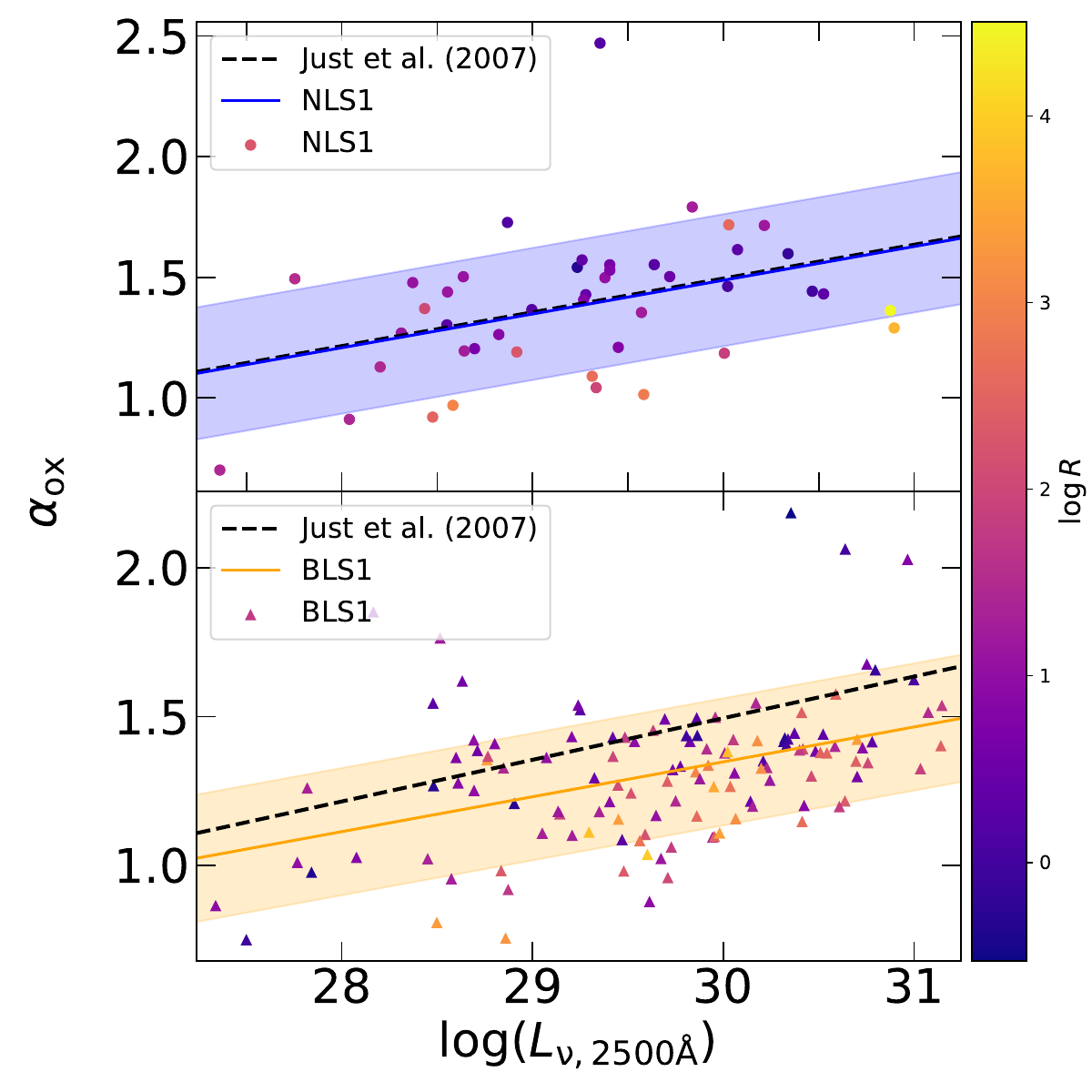}
	\caption{The correlation between $\alpha_{\rm ox}$ and $2500~\rm \AA$ luminosity for radio-detected NLS1s (upper panel) and BLS1s (lower panel). The coloured solid lines and the black dashed line represent the best-fit correlations for our samples and the \cite{2007ApJ...665.1004J} RQQs, respectively. The blue and orange-filled regions indicate the 1$  \sigma$ confidence intervals for the NLS1 and BLS1 sample data.\label{fig:aox-2500}}
\end{figure}

To further investigate the intrinsic differences in the optical/UV–X-ray slope $\alpha_{\rm ox}$ between NLS1s and BLS1s, we present in Fig. \ref{fig:aox-2500} the correlation of $\alpha_{\rm ox}$ with \( L_{\rm 2500\,\AA} \) for the two samples. Different colours are used to indicate the variation in radio loudness among individual sources. 
Both NLS1s and BLS1s show significant correlations between $\alpha_{\rm ox}$ and \( L_{\rm 2500\,\AA} \), with Spearman’s p-values less than 0.05. 
The blue and orange solid lines represent the best-fit correlations for the NLS1s and BLS1s, respectively: 
\[ \alpha_{\rm ox} = (0.140 \pm 0.051) \log(L_{\nu, 2500\,\rm \AA}) - (2.707 \pm 1.504) \quad (\text{NLS1s}), \] \[ \alpha_{\rm ox} = (0.117 \pm 0.023) \log(L_{\nu, 2500\,\rm \AA}) - (2.172 \pm 0.684) \quad (\text{BLS1s}). \]
For comparison, the black dashed line in Fig. \ref{fig:aox-2500} shows the two-parameter correlation for the bright RQQ sample of \cite{2007ApJ...665.1004J}. The slopes of our best-fitting for both samples are in excellent agreement with those of \cite{2007ApJ...665.1004J, 2010A&A...512A..34L}. Although our NLS1s have systematically lower optical luminosities than the \cite{2007ApJ...665.1004J} sample, the $\alpha_{\rm ox}$–$L_{\rm \nu,2500\AA}$ correlation remains nearly identical, as seen in the upper panel of Fig. \ref{fig:aox-2500}. In contrast, the majority of sources in the BLS1 sample lie below the best-fit line of the \cite{2007ApJ...665.1004J} RQQs in the $\alpha_{\rm ox}$–$L_{\rm \nu,2500\AA}$ plane. 

These results indicate that, for comparable optical luminosities, NLS1s and RQQs exhibit similar values of $\alpha_{\rm ox}$ \cite[in good agreement with the findings of][]{2023MNRAS.520.2781K}, implying similar X-ray luminosities. In contrast, BLS1s show smaller $\alpha_{\rm ox}$ values than RQQs, corresponding to stronger X-ray emission. 
This result is consistent with the statistical distribution of $\alpha_{\rm ox}$ in the sample, namely that matched-BLS1s with comparable optical luminosities have statistically lower $\alpha_{\rm ox}$ values than NLS1s. 

\subsubsection{$\alpha_{\rm ox}$ versus radio loudness}

\begin{figure}[htb!]
	\centering
	\includegraphics[width=0.4\textwidth]{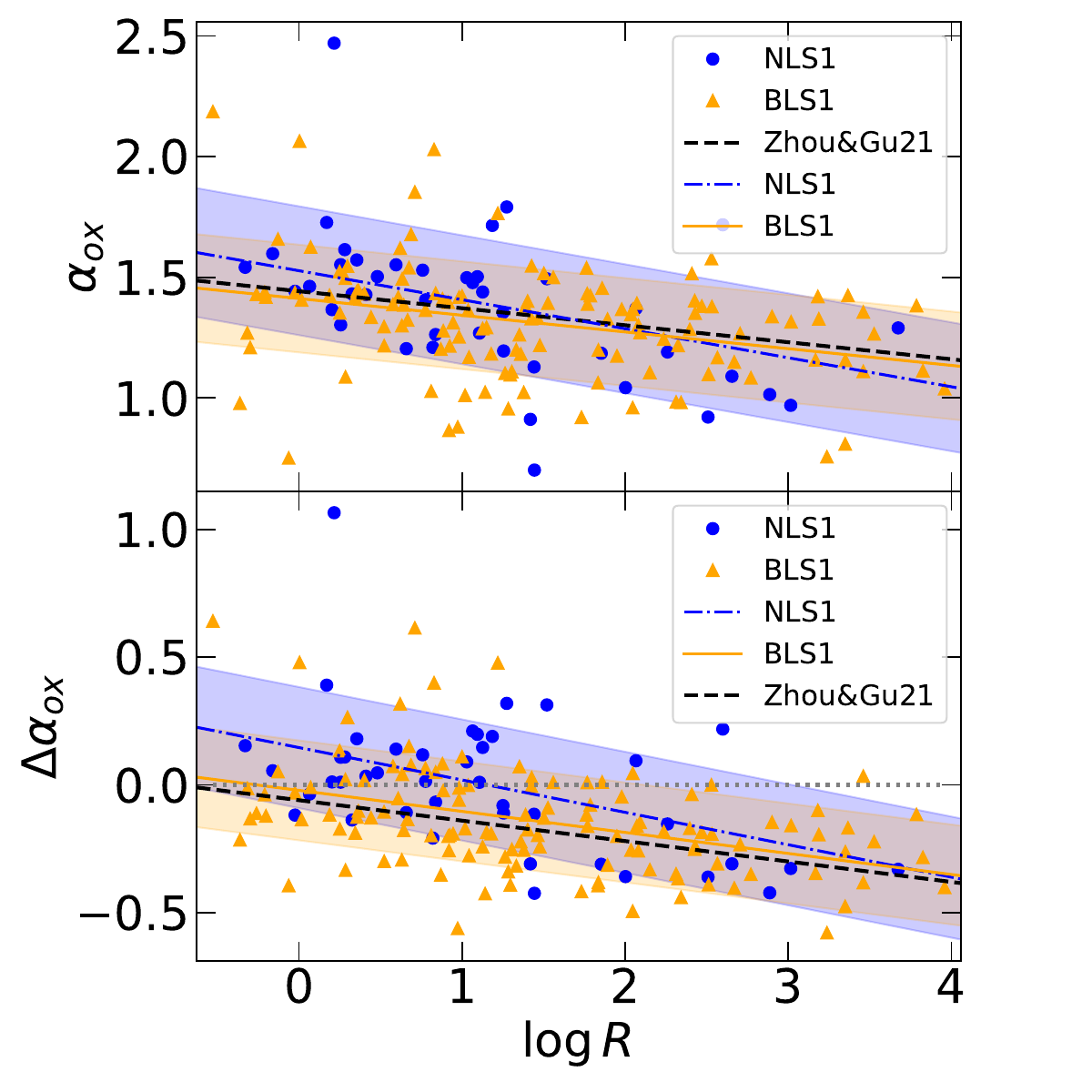}
	\caption{The relationship between radio loudness and $\alpha_{\rm ox}$ (upper panel) and $\Delta \alpha_{\rm ox}$ (lower panel) for radio-detected NLS1s (blue points) and BLS1s (orange triangles). The blue dotted, orange solid, and black dashed lines represent the best-fit correlations for the NLS1s, BLS1s, and the radio-detected quasar sample of \cite{2021RAA....21....4Z}, respectively. The blue and orange shaded regions denote the 1$\sigma$ confidence intervals for the NLS1 and BLS1 samples.
    \label{fig:aox-R}}
\end{figure}

The analysis of radio loudness in Fig. \ref{fig:aox-2500} is consistent with the results of \cite{2021RAA....21....4Z}: for NLS1s and BLS1s with higher radio loudness (warmer colours), their $\alpha_{\rm ox}$ values lie below the linear fit. That is, both samples exhibit that the higher the radio loudness, the smaller $\alpha_{\rm ox}$ at a fixed $L_{\rm \nu,2500\AA}$. This result is further confirmed in Fig. \ref{fig:aox-R}.
Fig. \ref{fig:aox-R} shows the correlations between radio loudness and $\alpha_{\rm ox}$, $\Delta \alpha_{\rm ox}$ for NLS1s and BLS1s. Here, $\Delta \alpha_{\rm ox} = \alpha_{\rm ox} - \alpha_{\rm ox,RQQ}$ represents the X-ray excess relative to radio-quiet quasars, where $\alpha_{\rm ox,RQQ}$ can be estimated from the $\alpha_{\rm ox}-L_{\rm \nu,2500\AA}$ relation in \cite{2007ApJ...665.1004J} using the $L_{\rm \nu,2500\AA}$ values of each source. 

Non-simultaneous optical and radio observations, which define the radio loudness, introduce additional scatter in the data presented in Fig. \ref{fig:aox-R}. Nevertheless, linear fitting of the correlation between $\alpha_{\rm ox}$ and radio loudness for each sample indicates that $\alpha_{\rm ox}$ is anti-correlated with radio loudness $\log R$ for both NLS1s and BLS1s (Spearman correlation $>99.9\%$). That is, the higher the radio loudness, the smaller the $\alpha_{\rm ox}$, and the stronger the X-ray emission normalized by optical luminosity. Our linear fitting results are: \[  \alpha_{\rm ox} = (-0.120 \pm 0.038) \log R + (1.527 \pm 0.061) \quad (\text{NLS1s}), \] \[  \alpha_{\rm ox} = (-0.069 \pm 0.019) \log R + (1.412 \pm 0.033) \quad (\text{BLS1s}). \] Compared with the results of \cite{2021RAA....21....4Z}, the dependence coefficient of $\alpha_{\rm ox}$ on $\log R$ is steeper for NLS1s, while the correlation distribution of $\alpha_{\rm ox}$ with $\log R$ for BLS1s is almost identical to that of \cite{2021RAA....21....4Z}.

In the bottom panel of Fig. \ref{fig:aox-R}, BLS1s exhibit a significant X-ray excess, with most points lying below the zero line, while NLS1s show a larger scatter. Most RQ-NLS1s (14/20) have $\Delta \alpha_{\rm ox} > 0$ with a mean value of 0.09, whereas RQ-BLS1s (33/51) predominantly show $\Delta \alpha_{\rm ox}<0$ with a mean value of $-0.04$. This suggests that, compared to typical radio-quiet AGNs \cite[][]{2007ApJ...665.1004J}, RQ-NLS1s do not exhibit a significant X-ray excess; instead, they show weaker X-ray emission. In contrast to RQ-NLS1s, RQ-BLS1s statistically exhibit $\Delta \alpha_{\rm ox}<0$, indicating a mild X-ray excess.
Nonetheless, the X-ray excess $\Delta \alpha_{\rm ox}$ is anti-correlated with radio loudness $\log R$ for both samples (Spearman correlation $>99.9\%$), i.e., the higher the radio loudness, the larger the X-ray excess. Our linear fitting results are: \[ \Delta \alpha_{\rm ox} = (-0.127 \pm 0.034) \log R + (0.146 \pm 0.054) \quad (\text{NLS1s}), \]  \[\Delta \alpha_{\rm ox} = (-0.082 \pm 0.016) \log R + (-0.021 \pm 0.029) \quad (\text{BLS1s}). \] Compared with the results of \cite{2021RAA....21....4Z}, the dependence coefficient of $\Delta \alpha_{\rm ox}$ on $\log R$ is steeper for NLS1s, while the correlation distribution of $\Delta \alpha_{\rm ox}$ with $\log R$ for BLS1s is almost identical to that of \cite{2021RAA....21....4Z}. 

\subsection{Fundamental Plane} 
\label{subsec:funda_plane}

\begin{figure*}[htb!]
	\sidecaption
	\includegraphics[width=6cm]{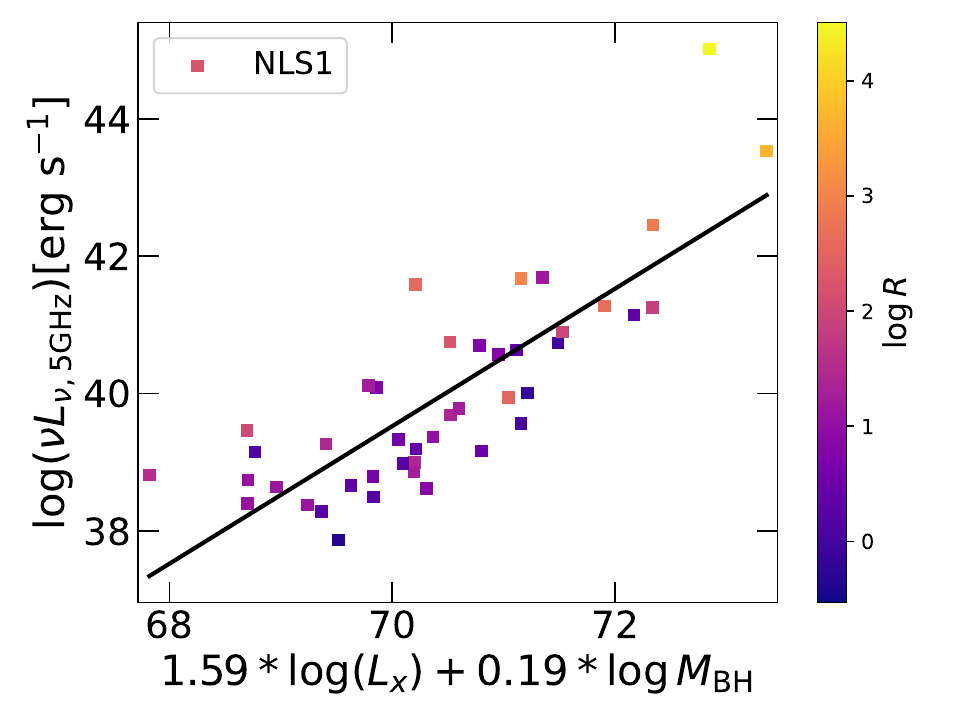}
    \includegraphics[width=6cm]{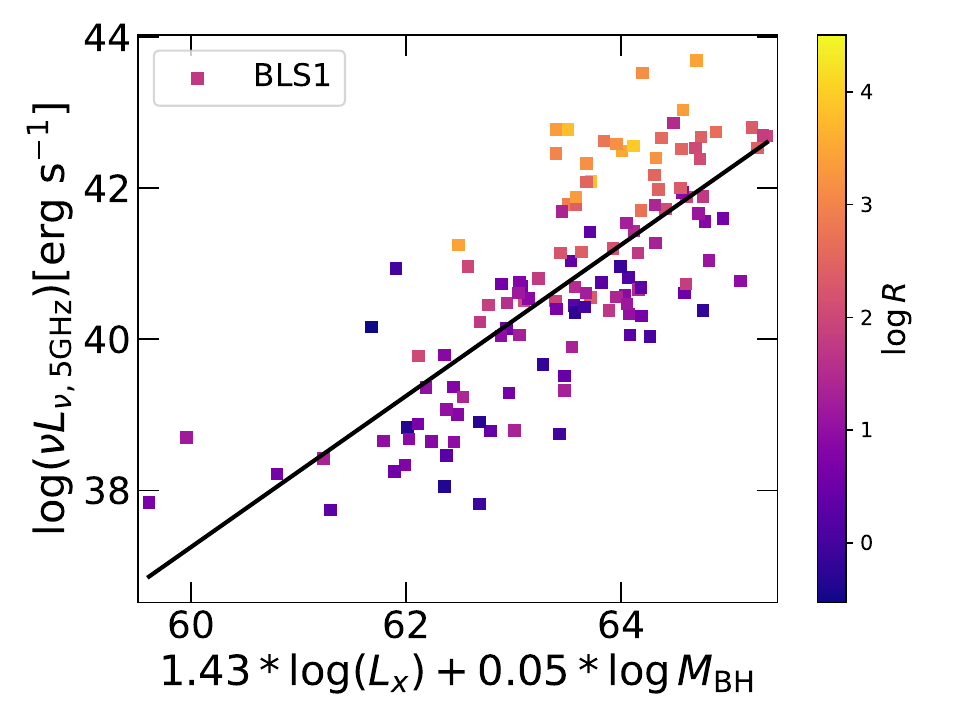}
	\caption{The dependence of radio luminosity on the X-ray luminosity and black hole mass (i.e. fundamental plane) in radio-detected NLS1s (Left panel) and BLS1s (Right panel).  \label{fig:L5-x-m}}
\end{figure*}

The fundamental plane of black hole accretion (the correlation among $5~\rm GHz$ radio luminosity, $2-10~\rm keV$ X-ray luminosity, and black hole mass) serves as an important probe for investigating the origin of X-ray emission \cite[e.g.,][]{2003MNRAS.345.1057M,2004A&A...414..895F,2006ApJ...645..890W,2008ApJ...688..826L,2018MNRAS.481L..45L}. Based on the multiple linear fit by the Markov chain Monte Carlo (MCMC) method with the Python-based $emcee$ code \cite[][]{2013PASP..125..306F}, the left and right panels of Fig. \ref{fig:L5-x-m} show the fundamental plane relations for NLS1s and BLS1s, respectively: 

{\tiny 
\[\log L_{\rm r} = (1.59\pm 0.32) \log L_{\rm X} + (0.19\pm 0.39) \log M_{\rm BH} - (30.47\pm 12.30) \quad (NLS1s),\] \[\log L_{\rm r} = (1.43\pm 0.19) \log L_{\rm X} + (0.05\pm 0.21) \log M_{\rm BH} - (22.75\pm 7.00) \quad (BLS1s).\]
}

These fundamental plane relations indicate that the radio luminosity of both NLS1s and BLS1s is strongly dependent on the X-ray luminosity. 
Although MCMC fitting suggests a slightly stronger black hole mass–radio luminosity dependence in NLS1s than in BLS1s, the correlation remains weak in both samples, and the difference is not statistically significant after accounting for the large uncertainties. 
Considering radio loudness, both samples exhibit a correlation between radio luminosity and radio loudness, as shown in Fig. \ref{fig:L5-x-m}: the larger the radio loudness (warmer colours), the more the distribution lies toward the upper right.

\subsection{Radio$-$X-ray relation} 
\label{subsec:r-x}

\begin{figure}[htb!]
	\centering
	\includegraphics[width=0.4\textwidth]{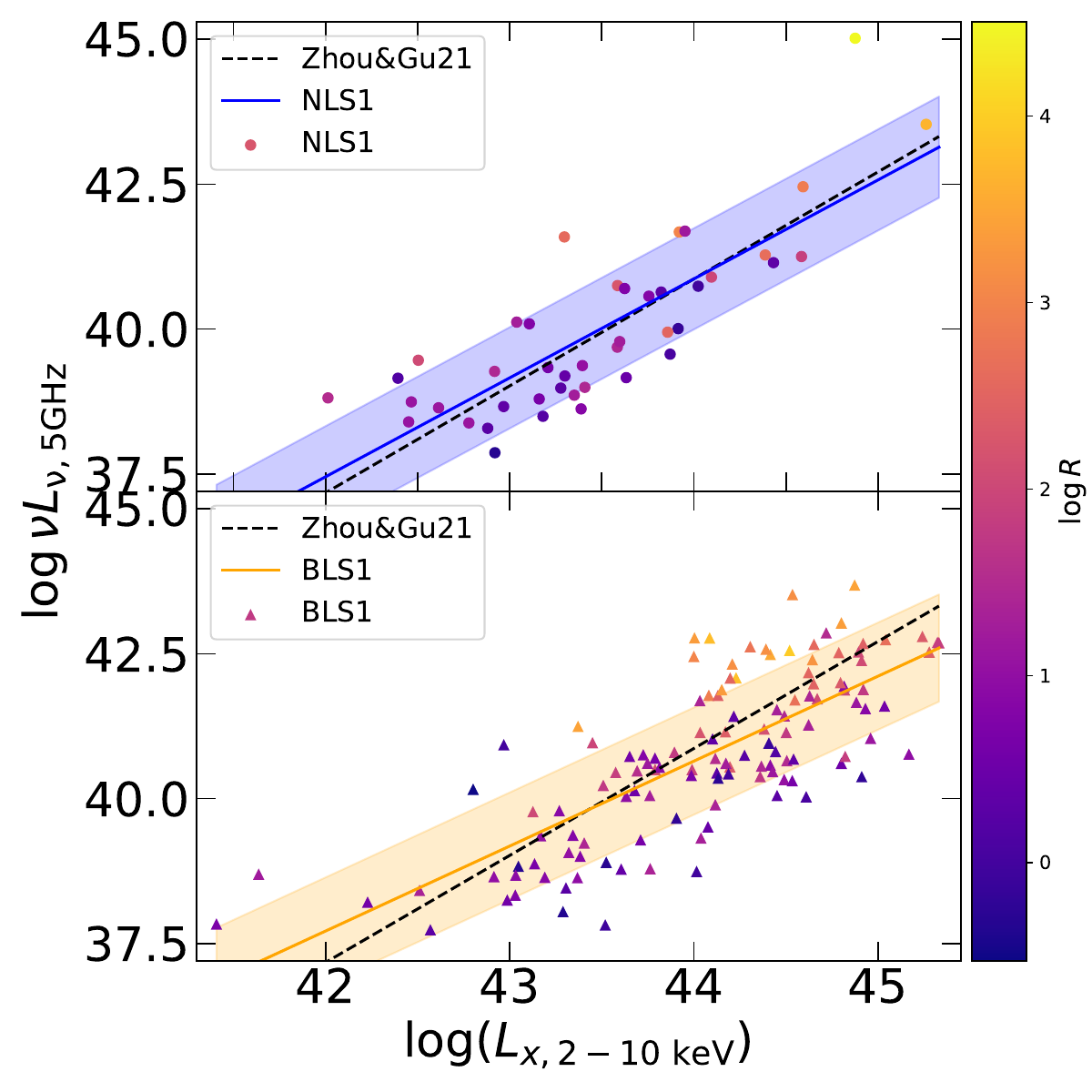}
	\caption{The relationship between radio and X-ray luminosity for radio-detected NLS1s (Upper panel) and BLS1s (Botom panel).  \label{fig:L5-x}}
\end{figure}

Given the strong correlation between radio and X-ray luminosities observed in the fundamental plane relations of NLS1s and BLS1s, we present the relationship between these two parameters for radio-detected quasars in Fig. \ref{fig:L5-x} and find a significant correlation for both NLS1s and BLS1s. The OLS bisector linear fit yields 
\[\log L_{\rm r} = (1.71\pm 0.19) \log L_{\rm X} - (34.25\pm 8.34) \quad (NLS1s),\] \[\log L_{\rm r} = (1.47\pm 0.11) \log L_{\rm X} - (23.85\pm 4.99) \quad (BLS1s).\]
Comparing the radio and X-ray correlations of RLQs detected by \cite{2021RAA....21....4Z}, as shown by the black dashed lines in Fig. \ref{fig:L5-x}, both RD-NLS1s and BLS1s exhibit a correlation similar to that of \cite{2021RAA....21....4Z} within the error range. 

\begin{figure*}[htb!]
	\sidecaption
	\includegraphics[width=6cm]{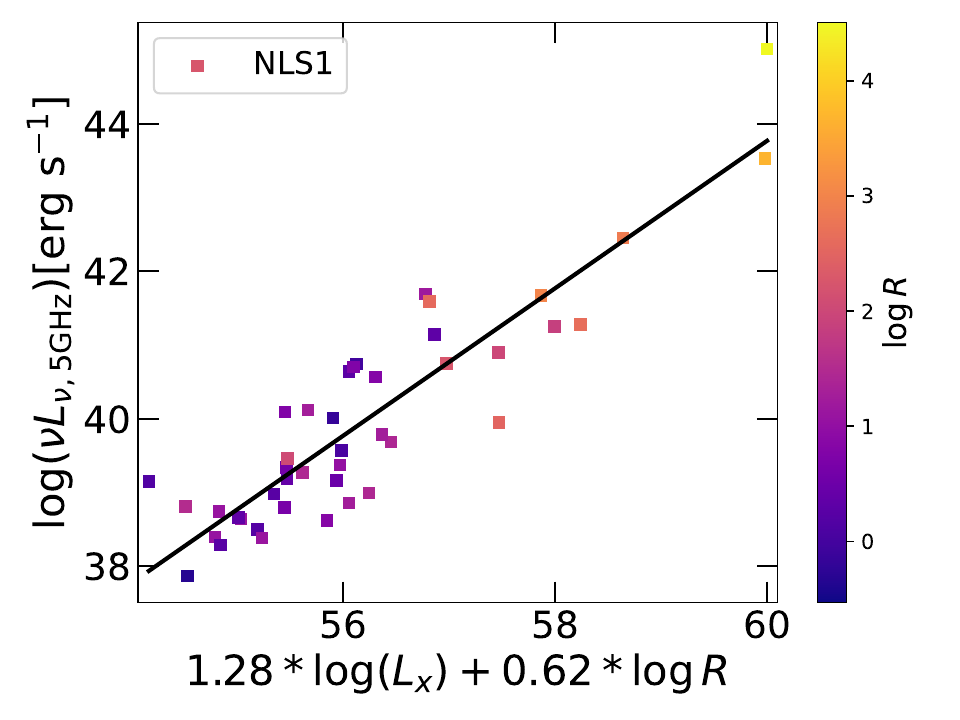}
    \includegraphics[width=6cm]{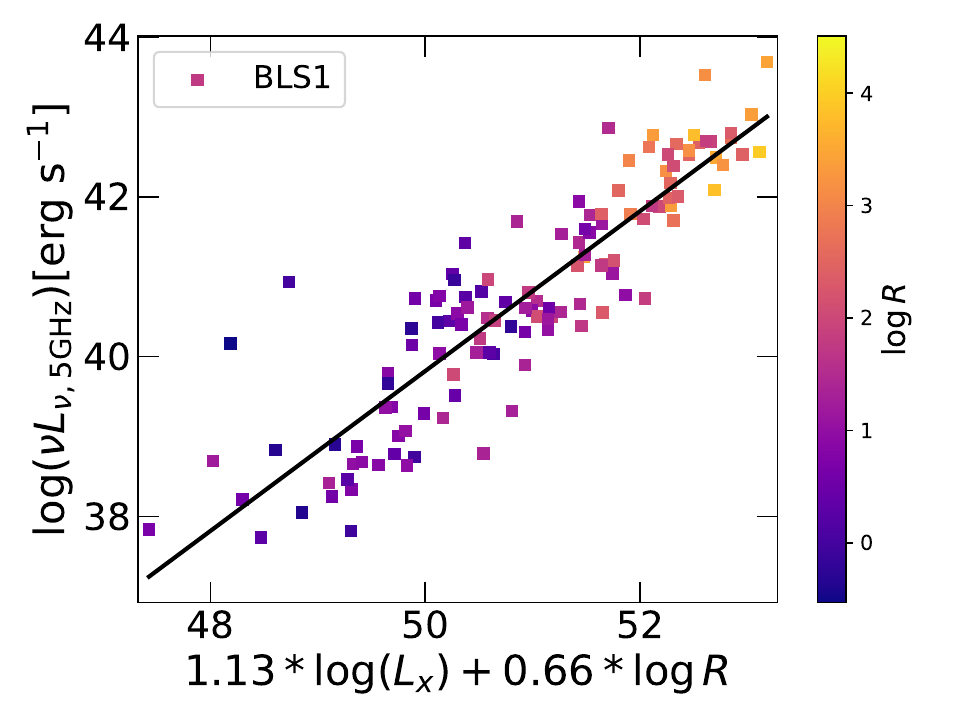}
	\caption{The dependence of radio luminosity on the X-ray luminosity and radio loudness in radio-detected NLS1s (Left panel) and BLS1s (Right panel).  \label{fig:L5-x-R}}
\end{figure*}

Similar to the fundamental plane relations for the two samples shown in Fig. \ref{fig:L5-x-m}, Fig. \ref{fig:L5-x} also reveals a dependence of radio luminosity on radio loudness. Considering radio loudness, we performed a multiple linear fit using $emcee$ for the relationship among the $5~\rm GHz$ radio luminosity, $2-10~\rm keV$ X-ray luminosity, and radio loudness $\log R$ for NLS1s and BLS1s. The results are presented in Fig. \ref{fig:L5-x-R}: 

{\tiny
\[ \log L_{\rm r} = (1.28\pm 0.25) \log L_{\rm X} + (0.62\pm 0.16) \log R - (16.22\pm 10.65) \quad (NLS1s),\] \[ \log L_{\rm r} = (1.13\pm 0.13) \log L_{\rm X} + (0.66\pm 0.09) \log R - (10.18\pm 5.69) \quad (BLS1s).\]
}
NLS1s and BLS1s exhibit similar correlations among radio luminosity, X-ray luminosity, and radio loudness. 
Comparing the results of \cite{2021RAA....21....4Z} for radio-detected quasars analyzed with the same method, our samples show a similar radio–X-ray correlation, but a slightly lower coefficient for the radio-luminosity–radio-loudness relation.

\section{Discussions} 
\label{sec:discc}

\subsection{Biases}

Although we have matched the NLS1 and BLS1 samples one-to-one in order to mitigate the impact of differences in their redshift and luminosity distributions on the statistical results related to X-ray luminosity, several inherent statistical biases remain, such as the non-simultaneous nature of multi-band observations, and sample completeness.

The radio, optical, and X-ray data used in this work were obtained from different instruments at non-simultaneous epochs. Although we excluded strongly variable sources during sample selection (Section \ref{sec:sample}), the intrinsic multi-band variability of AGNs inevitably introduces additional scatter into our multi-band correlation analyses, thereby affecting derived parameters such as radio loudness and the optical-to-X-ray slope $\alpha_{\rm ox}$. It is worth noting that parameters derived from multi-band observations do not characterize the instantaneous properties of a source; for example, radio loudness does not precisely correspond to jet power. Nevertheless, from a statistical perspective, these parameters collectively reflect the overall correlation trends within the sample.

This work investigates the X-ray emission properties of NLS1s and BLS1s based on X-ray spectral fitting. However, not all sources with X-ray observations were included in the spectral fitting, so the completeness of the X-ray data may affect the results to some extent. Through multi-wavelength data matching, we obtained 60 NLS1s and 163 BLS1s with optical, radio, and X-ray data. After X-ray spectral fitting, only 44 NLS1s and 125 BLS1s were retained for subsequent statistical comparisons. The X-ray completeness of the NLS1 and BLS1 samples is comparable (73.3\% and 76.7\%, respectively), indicating that this difference has only a minimal impact on the comparative results. 
Nevertheless, the numerous selection criteria leave us with a rather small final sample, with only 20 RQ-NLS1s after classification by radio loudness, which inevitably limits the statistical weight of our results. 

\subsection{Eddington Ratio}

\begin{figure}[htb!]
	\centering
	\includegraphics[width=0.4\textwidth]{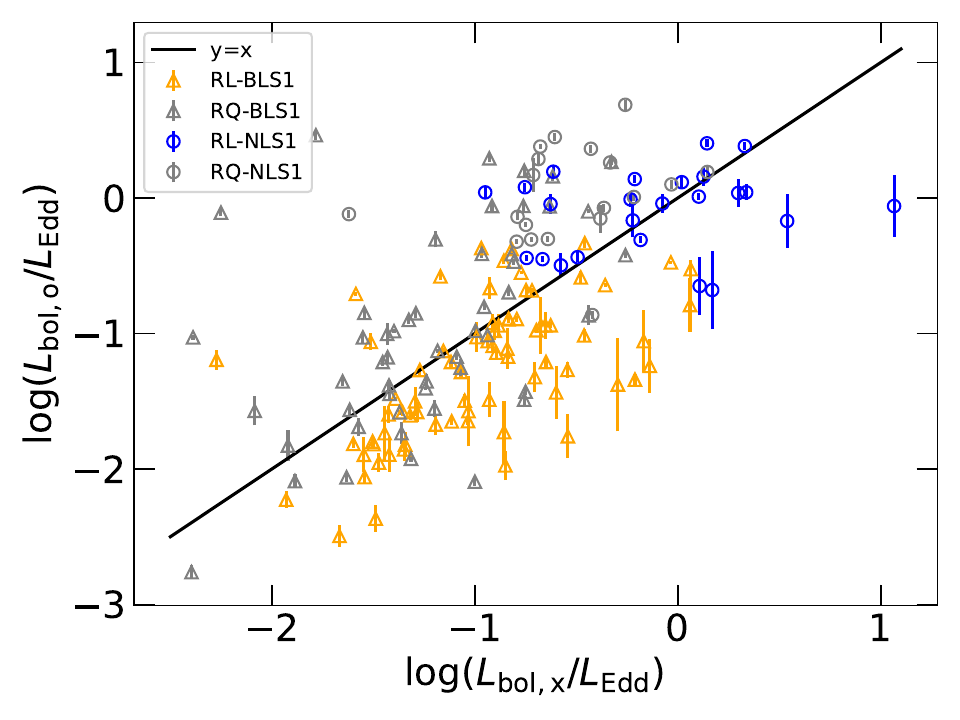}
    \caption{The X-ray to optical Eddington ratio correlations of radio-detected NLS1s (circles) and BLS1s (triangles).  \label{fig:R-x-o}}
\end{figure}

Within the framework of the AGN fundamental plane \citep{2003MNRAS.345.1057M} and the Eigenvector 1 main sequence quasars \citep{1992ApJS...80..109B}, the dimensionless Eddington ratio ($\lambda_{\rm Edd}=L_{\rm bol}/L_{\rm Edd}=L_{\rm bol}/(1.38\times 10^{38}M_{\rm BH}/M_\odot)$) serves as the primary driving parameter of the central engine. It governs the accretion disk–corona coupling, the amplitude of the soft X-ray excess, the morphology of broad emission lines, and the formation of jets, and it is the fundamental origin of the multi-wavelength observational differences between NLS1s and BLS1s \citep{2003A&A...398..927W}. 
Fig. \ref{fig:R-x-o} compares the Eddington ratios estimated from optical luminosity \cite[taken from][]{2024MNRAS.527.7055P} with those derived from X-ray luminosity in our sample. For the X-ray-based Eddington ratio, the bolometric luminosity is calculated from the $2-10~\rm keV$ luminosity using $L_{\rm bol,x} = K_{\rm bol} \times L_{2-10\rm keV}$, with $K_{\rm bol} = 15.86$ \citep{2024A&A...691A.203G}. 

When comparing the four subsamples of NLS1s and BLS1s classified by radio loudness in Fig. \ref{fig:R-x-o}, only the optical- and X-ray-based Eddington ratios of RQ-BLS1s are distributed near the equality line, consistent with the unobscured AGN sample of \cite{2024A&A...691A.203G}. The scatter among RQ-BLS1s may be attributed to their variability. In contrast to RQ-BLS1s, most RL-BLS1s exhibit significantly higher X-ray-based Eddington ratios than optical-based ones, suggesting the presence of an additional X-ray component in RL-BLS1s. Unlike RQ-BLS1s, the X-ray Eddington ratios of RQ-NLS1s are significantly lower than their optical counterparts, placing them above the equality line in Fig. \ref{fig:R-x-o}. Fig. \ref{fig:R-x-o} essentially shows the correlation between optical and X-ray bolometric luminosities, which rules out potential underestimation of central black hole masses in NLS1s \cite[e.g.,][]{2019ApJ...881L..24V}. Nevertheless, RQ-NLS1s and RQ-BLS1s still exhibit distinct distributions, indicating intrinsic differences in their X-ray emission. 
Similar to the case of RL-BLS1s relative to RQ-BLS1s, RL-NLS1s display stronger X-ray emission than RQ-NLS1s (see also in Fig. \ref{fig:aox-R}), causing the optical- and X-ray-based Eddington ratios of RL-NLS1s to lie close to the $y=x$ line. 

In the following text, we use optical-based Eddington ratio for correlation analysis.

\subsection{Soft X-ray emission}

Soft X-ray excess is a defining characteristic of NLS1s. Compared to BLS1s, NLS1s exhibit softer X-ray spectra \cite[][]{1996A&A...305...53B, 1999ApJS..125..317L, 2020MNRAS.498.5207W}. Recently, \cite{2020MNRAS.491..532G} systematically compared the soft X-ray emission properties of NLS1s and BLS1s, revealing that soft X-ray excess is common among Type 1 AGNs, including both NLS1s and BLS1s, but its strength is significantly greater in NLS1s.
Using Suzaku observations of 22 narrow-line Seyfert 1 galaxies (NLS1s) and 47 broad-line Seyfert 1 galaxies (BLS1s), \cite{2020MNRAS.498.5207W} further confirmed that NLS1s exhibit steeper power-law spectra and higher X-ray Eddington ratios. They also found that both the soft and hard X-ray excesses are stronger in NLS1s than in BLS1s, and that a correlation exists between the soft and hard excesses in NLS1s, whereas no such correlation is observed in BLS1s. The correlation between the soft and hard excesses in NLS1s may suggest a common origin for the two components, such as a blurred reflection model.
\cite{2022A&A...659A.172D} found that RQ-NLS1s exhibit a prominent soft X-ray excess, but unlike in BLS1s, their strength shows no significant positive correlation with the power-law photon index $\Gamma$, and they proposed that the excess could be attributed to a relativistic disk reflection model. 

Unlike previous studies that focused on samples dominated by radio-quiet NLS1s, we applied the same method to detect soft X-ray excesses and found that the soft X-ray excess detection rates for RD-NLS1s and BLS1s were not 100\%. 
Overall, the detection rate for NLS1s was significantly higher than that for BLS1s. After further classification by radio-loudness, the soft X-ray excess detection rates for radio-quiet NLS1s and BLS1s were each significantly higher than those for their respective radio-loud counterparts. 
The difference in soft X-ray excess detection rates may arise from the relative luminosities of the soft and hard X-ray components. Since the blackbody temperature distributions are comparable between the two samples, our results suggest that BLS1s have a higher power-law flux than NLS1s, consistent with the difference in their $\alpha_{\rm ox}$ distributions. Moreover, both samples exhibit differences in soft X-ray excess detection rates among different radio-loudness classes, supporting the idea that AGNs with stronger radio emission tend to have higher power-law X-ray luminosities, possibly due to jet activity \cite[e.g.,][]{2020MNRAS.496.2213D, 2021RAA....21....4Z}.

\begin{figure}[htb!]
	\centering
	\includegraphics[width=0.4\textwidth]{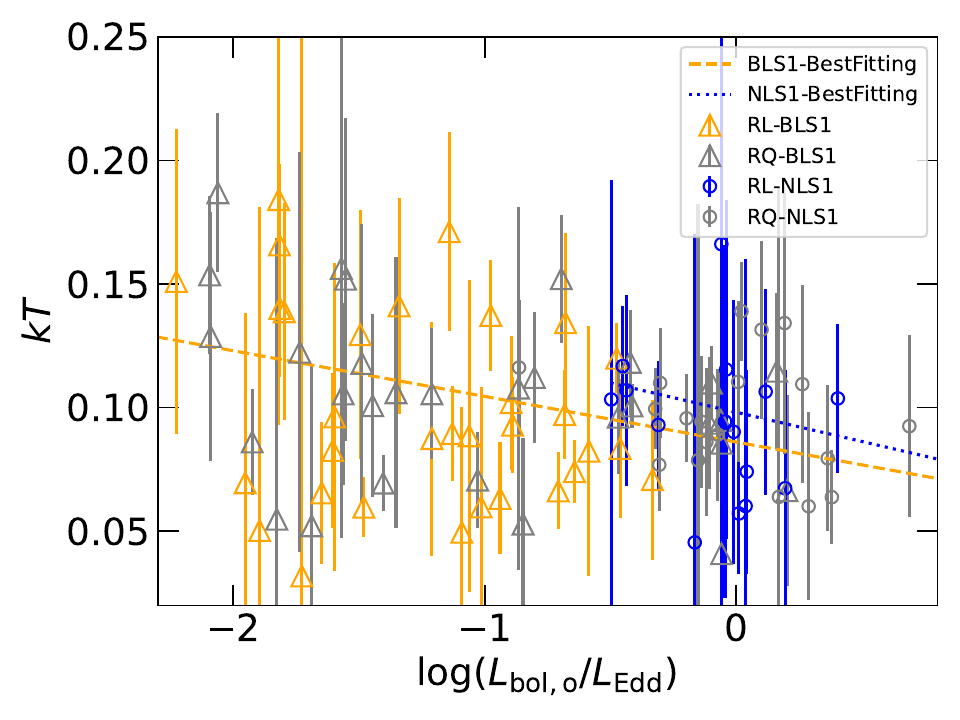}
	\caption{The relationship between blackbody temperature and optical-based Eddington ratios for radio-detected NLS1s (circles) and BLS1s (triangles).  \label{fig:kT-eddo}}
\end{figure}

Fig. \ref{fig:kT-eddo} compares the soft X-ray excess fitting temperatures with the Eddington ratio for NLS1s and BLS1s. Both samples exhibit similar trends in blackbody temperature as a function of the Eddington ratio. Despite the substantial scatter in Fig. \ref{fig:kT-eddo} for both samples—which yields a weak correlation for NLS1s (Spearman’s $p = 0.49$) but a significant one for BLS1s ($p = 0.004$)—the OLS linear fit slopes for NLS1s ($-0.024\pm0.013$) and BLS1s ($-0.018\pm0.006$) are consistent within errors. No significant difference is found between the radio-loud and radio-quiet subsamples, suggesting that radio emission does not significantly affect the soft X-ray excess component. 

There are three main models for the origin of the soft X-ray excess in AGNs\cite[e.g.,][]{2006MNRAS.365.1067C, 2020MNRAS.491..532G, 2022A&A...659A.172D, 2025ApJ...980...23C}: warm Comptonization, relativistic blurred reflection, and ionized gas absorption. Although spectral fitting of NLS1s and BLS1s reveals individual cases of warm absorption, this phenomenon is not common, and thus the ionized obscuration absorption model can be ruled out. The relativistic disk reflection model proposes that the soft X-ray excess originates from the reflection of hard X-rays off the accretion disk, implying a correlation between soft and hard X-ray emission. Our results show that RL-BLS1s and RL-NLS1s exhibit stronger $2~\rm keV$ emission (Fig. \ref{fig:aox-R}) and higher hard X-ray luminosities (Fig. \ref{fig:R-x-o}) compared to their radio-quiet counterparts. However, Fig. \ref{fig:kT-eddo} indicates no significant difference in blackbody temperature between the radio-loud and radio-quiet subsamples for either BLS1s or NLS1s. We further performed Spearman correlation tests between the soft X-ray excess temperature and the photon index as well as the $2-10~\rm keV$ luminosity for the radio-loud and radio-quiet subsamples in both samples. The resulting $p$-values are all much larger than 0.05, indicating no significant correlation between the soft X-ray excess temperature and hard X-ray emission across the radio-classified subsamples. These findings are inconsistent with the predictions of the relativistic disk reflection model. 
On the other hand, the soft X-ray supersoft blackbody temperature of BLS1s is significantly correlated with the Eddington ratio, supporting the warm Comptonization model, in which higher accretion rates lead to more efficient cooling, resulting in a lower warm corona temperature. In contrast, for NLS1s, although the fitted slope of the blackbody temperature–Eddington ratio correlation is similar to that of the BLS1s sample, the correlation is not significant enough to fully support the model. The details require further discussion with larger samples in the future.

\subsection{Hard X-ray emission}
\label{subsec:HardXray}

This study compares the X-ray emission of NLS1s and BLS1s using the optical/UV–X-ray spectral slope $\alpha_{\rm ox}$. Redshift- and luminosity-matched BLS1s and NLS1s exhibit slightly different $\alpha_{\rm ox}$ distributions, with BLS1s showing a lower median value. The correlation between $\alpha_{\rm ox}$ and UV luminosity (Fig. \ref{fig:aox-2500}) further indicates that BLS1s generally have lower $\alpha_{\rm ox}$ values (i.e., stronger X-ray emission) than NLS1s at the same optical/UV luminosity. Consistent with \cite{2021RAA....21....4Z}, our RD-NLS1s and BLS1s show strong correlations between X-ray emission and radio loudness. When comparing sub-samples based on radio loudness, the RQ-NLS1s exhibit weaker X-ray emission than RQ-BLS1s and do not show a significant X-ray excess. In Fig. \ref{fig:R-x-o}, which compares optical- and X-ray-based Eddington ratios (excluding possible effects of differences in black hole mass), RQ-NLS1s deviate from the line of equality between X-ray-based and optical-based Eddington ratios, unlike the RQ-BLS1s sample. In the fundamental relations, NLS1s and BLS1s show similar correlations between radio and X-ray luminosities, but show slightly different dependencies of radio luminosity on black hole mass, albeit with large uncertainties. Our results suggest possible physical differences in X-ray emission between RQ-NLS1s and RQ-BLS1s.

\begin{figure*}[htb!]
	\sidecaption
	\includegraphics[width=6cm]{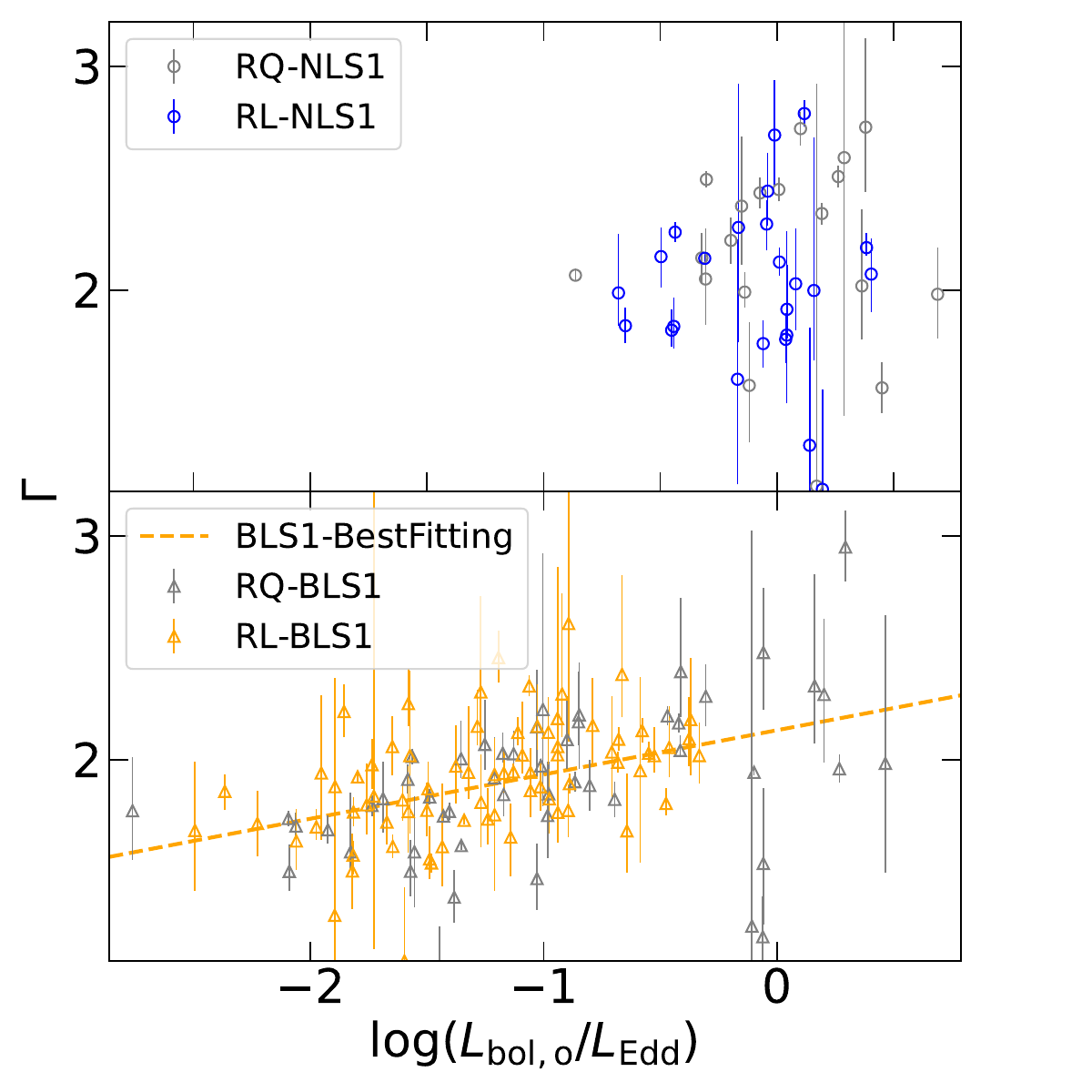}
        \includegraphics[width=6cm]{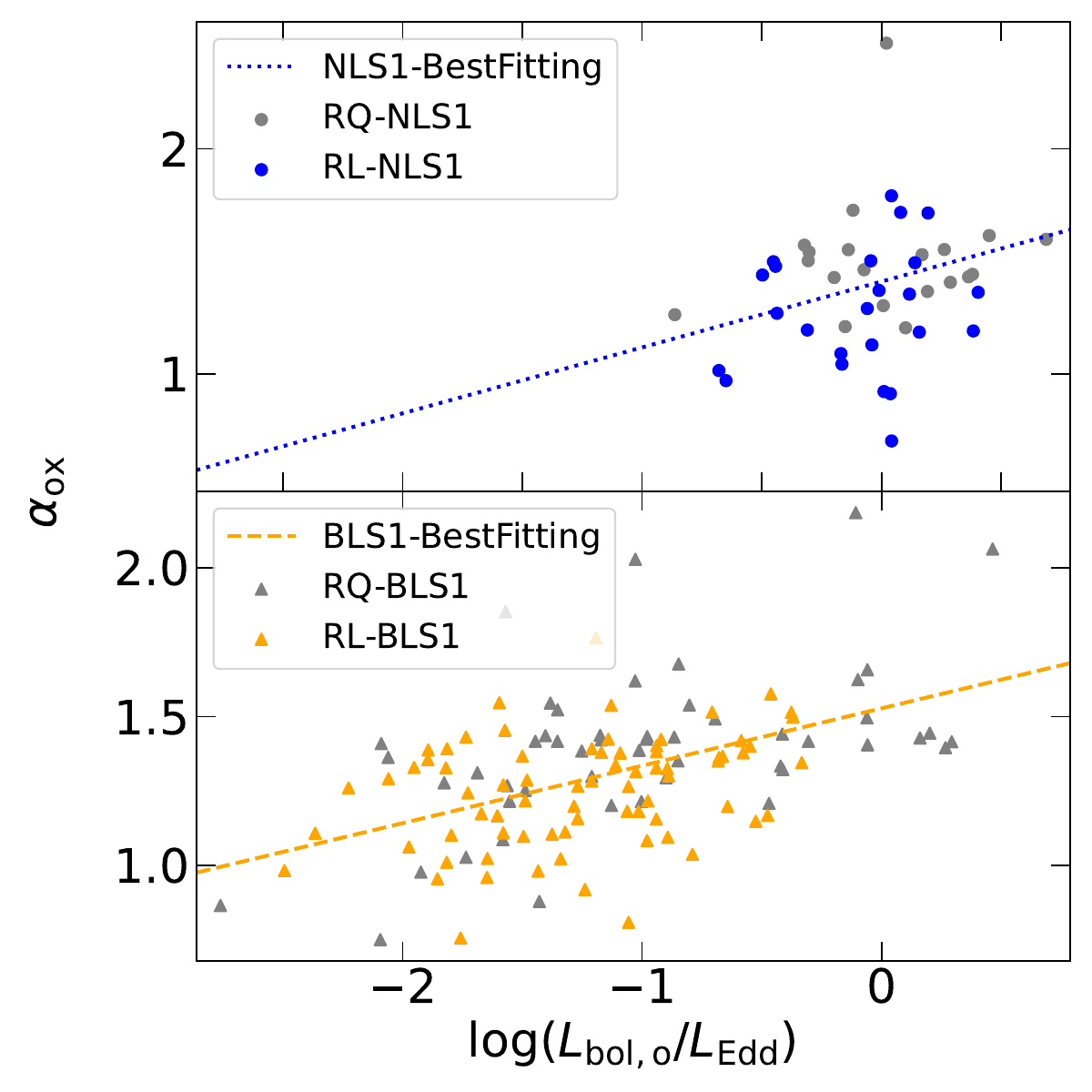}
	\caption{The left panels show the relationship between the X-ray photon index $\Gamma$ and the Eddington ratio, while the right panels show that between the optical/UV–X-ray slope $\alpha_{\rm ox}$ and the Eddington ratio, for radio-detected NLS1s (upper panels) and BLS1s (bottom panels).   \label{fig:gamma-eddo}}
\end{figure*}

It is generally accepted that the primary X-ray emission of RQ-AGN originates from inverse Comptonization in a hot corona, with a significant correlation between the hard X-ray photon index and the Eddington ratio \cite[e.g.,][]{2015A&ARv..23....1B, 2019MNRAS.490.3793L}. In contrast, RL-AGN do not show the correlation between the photon index and Eddington ratio \cite[e.g.,][]{2019MNRAS.490.3793L, 2020ApJ...893...39Z, 2020MNRAS.497..482L}. \cite{2020ApJ...896...95O} found that both NLS1s and BLS1s exhibit a correlation between the photon index and Eddington ratio, supporting the scenario that the X-ray emission in both samples originates from the AGN disk–corona model. Fig. \ref{fig:gamma-eddo} left panels compare the correlation between the photon index and Eddington ratio for our RD-NLS1s and BLS1s, with different radio-loudness sub-samples distinguished by different colours and symbols. The results show that BLS1s exhibit a significant positive correlation between the photon index and Eddington ratio (Spearman's $p = 2.36\times10^{-7}$). The ordinary least squares (OLS) bisector linear fit yields \( \Gamma = (0.16\pm0.04)\log (L_{\rm bol,o}/L_{\rm Edd}) + (2.12\pm0.05) \), consistent with the results of \cite{2020ApJ...896...95O}, but with flatter slop than that of RQ-AGNs in \cite[with slop about 0.31][]{2008ApJ...682...81S, 2019MNRAS.490.3793L}, and similar correlations are found for both RL-BLS1s and RQ-BLS1s. However, contrary to the findings of \cite{2020ApJ...896...95O}, NLS1s do not show a significant correlation between the photon index and Eddington ratio (Spearman’s $p$-values of 0.34), and the same holds for the two NLS1s sub-samples classified by radio loudness. 

The right panels of Fig. \ref{fig:gamma-eddo} compare the $\alpha_{\rm ox}$–Eddington ratio correlation for RD-NLS1s and BLS1s. Same with previous results \cite[e.g.,][]{2010ApJS..187...64G, 2024A&A...691A.203G}, the RL-BLS1s, RQ-BLS1s, and the total BLS1s sample all exhibit a consistent $\alpha_{\rm ox}$–Eddington ratio correlation, with Spearman's $p = 3.05\times10^{-9}$. In contrast, the correlation for NLS1s is weak, with $p = 0.04$.  

In the disk–corona model of radio-quiet active galactic nuclei (RQ-AGN), a higher Eddington ratio corresponds to stronger Compton cooling, resulting in steeper $2-10~\rm keV$ power-law spectra. The positive correlations between the $\Gamma-$, $\alpha_{\rm ox}-$ and Eddington ratio observed in our BLS1s sample support a disk–corona contribution for their X-ray emission. However, the slope of this correlation is flatter than that typically found in RQ-AGN, which may indicate a slower coronal cooling rate in BLS1s or the presence of an additional X-ray component associated with radio jets. 
On the other hand, compared with normal NLS1s \cite[][]{2011ApJ...727...31A, 2020MNRAS.491..532G, 2020MNRAS.498.5207W, 2020ApJ...896...95O}, the lack of a significant correlation between $\Gamma$ and the Eddington ratio in our RD-NLS1s sample suggests the possible intrinsic physical differences in X-ray emission between NLS1s and BLS1s. 
This apparent discrepancy can be comprehensively explained by two fundamental differences in sample composition and physical contamination.
First, unlike previous studies dominated by pure accretion-driven sources (RQ-NLS1s), our entire NLS1 sample is strictly radio-detected, containing both radio-loud and faint radio-core sources. Even RQ-NLS1s selected from radio surveys may be harbor weak compact nuclear jets or persistent jet-base non-thermal emission \cite[e.g.,][]{2020A&A...636A..64B, 2025ApJ...987L..26W}. 
This additional jet-related hard X-ray component contaminates the intrinsic coronal spectrum, flattens the overall photon index, and introduces substantial source-to-source scatter that washes out the clean accretion-driven $\Gamma$–$\lambda_{\rm Edd}$ trend. 
Second, jet variability further decouples spectral shape from accretion state. RD-NLS1s commonly exhibit intermittent jet activity, whose variable non-thermal hard X-ray flux distorts the steady corona scaling established in purely RQ-NLS1 samples. 

Admittedly, owing to the limited sample size (only 44 RD-NLS1s), our results are subject to some statistical scatter. In the future, incorporating more X-ray data from facilities such as Chandra and eROSITA may allow us to further investigate the differences in X-ray emission between RD-NLS1s and BLS1s.

\subsection{Radio Loudness}

Consistent with the findings that most radio-loud AGN exhibit stronger X-ray emission compared to radio-quiet samples\cite[][]{2011ApJ...726...20M,2020MNRAS.496..245Z,2021RAA....21....4Z,2023MNRAS.520.2781K}, both RL-NLS1s and RL-BLS1s show higher X-ray radiation than their radio-quiet counterparts at similar ultraviolet luminosities. Regarding the extra jet component in the X-ray emission of RL-AGN\cite[e.g.,][]{2011ApJ...726...20M}, \cite{2020MNRAS.496..245Z} argued that the strong X-ray radiation in RL-AGN is essentially a consequence of the disk-corona structure evolution, with the jet modulating the physical state of the corona to enhance its emission. Although the jet-regulated corona model proposed by \cite{2020MNRAS.496..245Z} can account for the effect of radio loudness on X-ray radiation, it does not adequately explain the difference in X-ray emission between NLS1s and BLS1s found in this work.

The accretion rates of both BLS1s and NLS1s in our sample are generally greater than 0.01, with no significant transition in accretion state, indicating similar accretion states, as evidenced by their identical $\Gamma-$ and $\alpha_{\rm ox}-$ Eddington ratio correlations \cite[e.g.,][]{2010ApJS..187...64G, 2020ApJ...896...95O}. In the same disk-corona model, RQ-NLS1s and RQ-BLS1s should have similar X-ray properties. However, Fig. \ref{fig:aox-R} reveals a discrepancy in \( \Delta \alpha_{\rm ox} \) between the two populations: most RQ-BLS1s have \( \Delta \alpha_{\rm ox}<0 \) (indicating an X-ray excess), while most RQ-NLS1s exhibit \( \Delta \alpha_{\rm ox}>0 \) (indicating an X-ray deficit). The correlation between X-ray Eddington ratio and optical Eddington ratio for the two samples in Fig. \ref{fig:R-x-o} further supports the difference in X-ray luminosity between RQ-NLS1s and RQ-BLS1s. For RQ-BLS1s, the X-ray Eddington ratio is comparable to the optical Eddington ratio, whereas for RQ-NLS1s, the X-ray Eddington ratio is statistically lower than the optical one, suggesting that the X-ray Eddington ratio estimated using the RQ-AGN scaling is underestimated for RQ-NLS1s, i.e., they have weaker X-ray emission. 

The extra-jet X-ray emission model can account for all the observed phenomena in this work. All NLS1s in our sample are radio-detected AGN and are considered to be in an early stage of jet evolution. 
For RQ-NLS1s in the early stage (jet formation phase), the coronal emission may be suppressed by the nascent jet, while the jet X-ray emission is also weak, resulting in even weaker overall X-ray emission. As the jet emission strengthens, the X-ray emission from NLS1s gradually increases. In contrast, radio-detected BLS1s are in a phase of stable accretion and stable jet, where the X-ray emission contains both coronal and jet components. Their X-ray excess \( \Delta \alpha_{\rm ox} \) shows a significant positive correlation with radio loudness. The correlations among radio luminosity, X-ray luminosity, and radio loudness in both samples further support the jet X-ray model. 

As discussed above, the definition of radio loudness in this paper is based on the radio emission from individual sources. However, the radio emission of AGN is not only to relativistic jets. In particular, for radio-quiet AGNs, contributions from outflows \cite[][]{2024MNRAS.528.3696L} and star formation-related radio emission \cite[e.g.,][]{2019NatAs...3..387P} cannot be neglected. Although radio loudness can statistically reflect the strength of AGN jet power, the limited sample size makes it difficult to fully separate radio-quiet AGN outflows, star formation, and other radio emissions. Future studies that expand the sample with Chandra and eROSITA data and perform pc-scale imaging with VLBA and SKA to investigate the X-ray properties of NLS1s classified according to jet presence will further validate the results presented here.

\section{Summary} \label{sec:sum}

To investigate the X-ray emission properties of early-stage accretion activity or early jets in NLS1s, this work presents the first large-sample comparative study of radio-detected NLS1s and BLS1s, based on FIRST, SDSS spectroscopy, and XMM-Newton observations, focusing on their X-ray emission and multi-band correlations. The main results are as follows:

(1) Through XMM-Newton X-ray spectral fitting, we find that NLS1s have a higher detection rate of soft X-ray excess compared to BLS1s. Although the soft X-ray excess detection rate of RL-NLS1s is lower than that of RQ-NLS1s, it remains higher than that of RL-BLS1s.

(2) Compared to BLS1s, NLS1s have comparable blackbody temperature, but exhibit a slightly steeper X-ray power-law photon index. 

(3) When matched in redshift and luminosity with BLS1s, NLS1s exhibit a slightly larger optical/UV-to-X-ray spectral index \( \alpha_{\rm ox} \), i.e., slightly weaker X-ray emission, albeit with larger scatter. 

(4) Both samples show a similar correlation between \( \alpha_{\rm ox} \) and UV luminosity: stronger UV luminosity corresponds to steeper \( \alpha_{\rm ox} \). However, at comparable UV luminosity, BLS1s have a smaller \( \alpha_{\rm ox} \), indicating stronger X-ray emission. 

(5) In both samples, \( \alpha_{\rm ox} \) is strongly correlated with radio loudness: higher radio loudness corresponds to smaller \( \alpha_{\rm ox} \), though the slopes differ slightly.

(6) The X-ray excess $\Delta \alpha_{\rm ox}$ differs between RQ-NLS1s and RQ-BLS1s: most RQ-NLS1s have $\Delta \alpha_{\rm ox}>0$, indicating an X-ray deficit, while RQ-BLS1s show the opposite, i.e., an X-ray excess. Overall, $\Delta \alpha_{\rm ox}$ in both samples is significantly correlated with radio loudness, with greater radio loudness corresponding to a more pronounced X-ray excess.

(7) The fundamental plane indicates that BLS1s show no correlation between radio luminosity and black hole mass, whereas NLS1s exhibit a weak correlation, albeit with large uncertainties. 

(8) The multivariate linear relationships among radio luminosity, X-ray luminosity, and radio loudness are similar for the two samples. 

(9) For RQ-NLS1s, the Eddington ratio calculated from X-ray luminosity is systematically lower than the optical-based Eddington ratio, confirming their intrinsic deficit of X-ray emission compared with RQ-BLS1s under identical accretion conditions.

In summary, radio-detected NLS1s and BLS1s exhibit intrinsic differences in X-ray emission. This likely stems from the fact that NLS1s are in a radio-young phase: the jet formation process may lead to weaker X-ray radiation in RQ-NLS1s. These findings are fully consistent with the jet X-ray emission model, which warrants further discussion in future work.


\begin{acknowledgements}
We thank the anonymous referee for valuable and insightful suggestions that improved the manuscript.
This work is supported by Jiangxi Provincial Natural Science Foundation (No. 20232BAB211024), the Doctoral Scientific Research Foundation of Shangrao Normal University (Grant No. K6000449), and the project of Jiangxi Province Key Laboratory of Applied Optical Technology (No. 2024SSY03051). 
MFG is supported by the National Science Foundation of China (grant 12473019), the China Manned Space Project with No. CMS-CSST-2025-A07, the National SKA Program of China (Grant No. 2022SKA0120102), and the Shanghai Pilot Program for Basic Research-Chinese Academy of Science, Shanghai Branch (JCYJ-SHFY-2021-013). 
ML is supported by the grants from the Rubin-Chile Fund (DIA3324). 
ZC is supported by the National Natural Science Foundation of China under grant Nos. 12303027. 
This work is based on results from the enhanced XMM-Newton spectral-fit database, an ESA PRODEX funded project, based in turn on observations obtained with XMM-Newton, an ESA science mission with instruments and contributions directly funded by ESA Member States and NASA. 
Funding for SDSS-III has been provided by the Alfred P. Sloan Foundation, the Participating Institutions, the National Science Foundation, and the U.S. Department of Energy Office of Science. The SDSS-III web site is http://www.sdss3.org/. 
\end{acknowledgements}

\begin{appendix}

\onecolumn
\begin{landscape}

\section{Spectral fitting results}
\label{appendix1}

\begin{table}[h]
\caption{Spectral fitting results for RD-NLS1s and RD-BLS1s}
\label{tableS2.tab} 
\centering
\begin{tabular}{c|cccccccccl}
\hline        
\multicolumn{1}{c}{Name} & \multicolumn{1}{c}{XMM.ID} & \multicolumn{1}{c}{z.nH} & \multicolumn{1}{c}{kT} & \multicolumn{1}{c}{bbd.norm} & \multicolumn{1}{c}{$\Gamma$} & \multicolumn{1}{c}{pow.norm} & \multicolumn{1}{c}{$\chi^2/d.o.f$} & \multicolumn{1}{c}{$\log F_{2-10\rm keV}$} & \multicolumn{1}{c}{$f_{2\rm keV}$} & \multicolumn{1}{c}{Note} \\
\multicolumn{1}{c}{SDSS J} & \multicolumn{1}{c}{ } & \multicolumn{1}{c}{$10^{22}~\rm cm^{-2}$} & \multicolumn{1}{c}{$\rm keV$} & \multicolumn{1}{c}{$\times 10^{-5}$} & \multicolumn{1}{c}{ } & \multicolumn{1}{c}{$\times 10^{-4}$} & \multicolumn{1}{c}{ } & \multicolumn{1}{c}{$\rm erg~s^{-1}~cm^{-2}$} & \multicolumn{1}{c}{$\rm erg~s^{-1}~cm^{-2}~Hz^{-1}$} & \multicolumn{1}{c}{ } \\
\multicolumn{1}{c}{(1)} & \multicolumn{1}{c}{(2)} & \multicolumn{1}{c}{(3)} & \multicolumn{1}{c}{(4)} & \multicolumn{1}{c}{(5)} & \multicolumn{1}{c}{(6)} & \multicolumn{1}{c}{(7)} & \multicolumn{1}{c}{(8)} & \multicolumn{1}{c}{(9)} & \multicolumn{1}{c}{(10)} & \multicolumn{1}{c}{(11)} \\
\hline
\multicolumn{11}{c}{NLS1 sample} \\ 
\hline
011448.67-002946.0 & 0747400132 & $<0.30$  &  $0.095_{-0.03}^{+0.04}$  & $0.21$ &  $1.31_{-0.36}^{+0.53}$ &  $0.51_{-0.17}^{+0.43} $  & 8.7/12 & $-12.406_{-0.055}^{+0.049}$  & 2.77$\rm E-$31 & 0 \\ 
013955.76+061922.5 & 0110890901 & $<0.001$  &  $0.093_{-0.003}^{+0.003}$  & 2.63 &  $1.98_{-0.20}^{+0.21}$ &  $0.56_{-0.09}^{+0.10} $  & 46.41/40 & $-12.827_{-0.031}^{+0.029}$  & 2.72$\rm E-$31 & 0 \\ 
030313.02-001457.5 & 0803952301 & $<0.001$  &  $0.166_{-0.020}^{+0.019}$  & 0.26 &  $1.76_{-0.11}^{+0.11}$ &  $2.23_{-0.25}^{+0.26} $  & 68.39/78 & $-12.083_{-0.014}^{+0.013}$  & 1.55$\rm E-$30 & 0 \\ 
\multicolumn{11}{c}{...} \\ 
\hline
\multicolumn{11}{c}{BLS1 sample} \\
\hline
000710.01+005329.0 & 0821730101 & $<0.001$  & $...$  & $...$ &  $1.91_{-0.06}^{+0.06}$  & $2.95_{-0.11}^{+0.11}$  & 71.96/76 & $-12.062_{-0.017}^{+0.016}$ & 1.33$\rm E-$30 & 0\\ 
001137.24+144201.4 & 0600090101 & $<0.001$  & $0.134^{+0.022}_{-0.023}$  & $0.157$ &  $2.09_{-0.05}^{+0.05}$  & $2.03_{-0.06}^{+0.06}$  & 77.87/77 & $-12.341_{-0.011}^{+0.010}$ & 7.25$\rm E-$31 & 0\\ 
004607.97+090720.9 & 0886221501 & $3.045_{-0.300}^{+0.644}$  & $...$  & $...$ &  $2.03_{-0.04}^{+0.09}$  & $1.92_{-0.10}^{+0.19}$  & 119.32/97 & $-12.325_{-0.010}^{+0.010}$ & 7.77$\rm E-$31 & zxipcf\\ 
\multicolumn{11}{c}{...} \\ 
\hline
\end{tabular}
\tablefoot{
In this Table, Column (1): SDSS name; Column (2): The XMM-Newton observation ID; Column (3): The intrinsic hydrogen column density; Column (4$ - $5): parameters of blackbody components; Column (6$ - $7): parameters of power-law components; Column(8): $\chi^2$ and degrees of freedom; Column ($9$): The powerlaw flux at $2-10\rm \, keV$; Column (10): The flux at rest frame $2\rm \, keV$; Column (11): The notes on the XMM-Newton X-ray spectral fitting process, such as absorption effects from zxipcf and zedge, or the effect of pile-up. Table \ref{tableS2.tab} is published in its entirety in the machine-readable format.}
\end{table}

\end{landscape}

\end{appendix}

\FloatBarrier 
\clearpage


\begin{thebibliography}{}

\bibitem[Abdo et al.(2009)]{2009ApJ...707L.142A} Abdo, A.~A., Ackermann, M., Ajello, M., et al.\ 2009, \apjl, 707, 2, L142
\bibitem[Ai et al.(2011)]{2011ApJ...727...31A} Ai, Y.~L., Yuan, W., Zhou, H.~Y., et al.\ 2011, \apj, 727, 1, 31
\bibitem[Antonucci(1993)]{1993ARA&A..31..473A} Antonucci, R.\ 1993, \araa, 31, 473
\bibitem[Berton et al.(2019)]{2019A&A...632A.120B} Berton, M., Braito, V., Mathur, S., et al.\ 2019, \aap, 632, A120
\bibitem[Berton et al.(2020)]{2020A&A...636A..64B} Berton, M., J{\"a}rvel{\"a}, E., Crepaldi, L., et al.\ 2020, \aap, 636, A64
\bibitem[Boller et al.(1996)]{1996A&A...305...53B} Boller, T., Brandt, W.~N., \& Fink, H.\ 1996, \aap, 305, 53
\bibitem[Boroson \& Green(1992)]{1992ApJS...80..109B} Boroson, T.~A. \& Green, R.~F.\ 1992, \apjs, 80, 109
\bibitem[Boroson(2002)]{2002ApJ...565...78B} Boroson, T.~A.\ 2002, \apj, 565, 1, 78
\bibitem[Brandt \& Alexander(2015)]{2015A&ARv..23....1B} Brandt, W.~N. \& Alexander, D.~M.\ 2015, \aapr, 23, 1
\bibitem[Caccianiga et al.(2014)]{2014MNRAS.441..172C} Caccianiga, A., Ant{\'o}n, S., Ballo, L., et al.\ 2014, \mnras, 441, 1, 172
\bibitem[Chaudhary et al.(2026)]{2026ApJ...996..118C} Chaudhary, S.~C., Prince, R., van Soelen, B., et al.\ 2026, \apj, 996, 2, 118
\bibitem[Chen et al.(2025)]{2025ApJ...980...23C} Chen, S.-J., Wang, J.-X., Kang, J.-L., et al.\ 2025, \apj, 980, 1, 23
\bibitem[Chen et al.(2025)]{2025ApJ...995..197C} Chen, S.-J., Wang, J.-X., Kang, J.-L., et al.\ 2025, \apj, 995, 2, 197
\bibitem[Chen et al.(2022)]{2022RAA....22i5006C} Chen, Y., Gu, Q., Fan, J., et al.\ 2022, Research in Astronomy and Astrophysics, 22, 9, 095006
\bibitem[Collin \& Kawaguchi(2004)]{2004A&A...426..797C} Collin, S. \& Kawaguchi, T.\ 2004, \aap, 426, 797
\bibitem[Crummy et al.(2006)]{2006MNRAS.365.1067C} Crummy, J., Fabian, A.~C., Gallo, L., et al.\ 2006, \mnras, 365, 4, 1067
\bibitem[D'Ammando(2020)]{2020MNRAS.496.2213D} D'Ammando, F.\ 2020, \mnras, 496, 2, 2213
\bibitem[Dehghanian et al.(2019)]{2019ApJ...882L..30D} Dehghanian, M., Ferland, G.~J., Peterson, B.~M., et al.\ 2019, \apjl, 882, 2, L30
\bibitem[Ding et al.(2022)]{2022A&A...659A.172D} Ding, N., Gu, Q., Tang, Y., et al.\ 2022, \aap, 659, A172
\bibitem[Fabian et al.(2004)]{2004MNRAS.353.1071F} Fabian, A.~C., Miniutti, G., Gallo, L., et al.\ 2004, \mnras, 353, 4, 1071
\bibitem[Falcke et al.(2004)]{2004A&A...414..895F} Falcke, H., K{\"o}rding, E., \& Markoff, S.\ 2004, \aap, 414, 895
\bibitem[Foreman-Mackey et al.(2013)]{2013PASP..125..306F} Foreman-Mackey, D., Hogg, D.~W., Lang, D., et al.\ 2013, \pasp, 125, 925, 306
\bibitem[Foschini et al.(2015)]{2015A&A...575A..13F} Foschini, L., Berton, M., Caccianiga, A., et al.\ 2015, \aap, 575, A13
\bibitem[Ghosh et al.(2018)]{2018MNRAS.479.2464G} Ghosh, R., Dewangan, G.~C., Mallick, L., et al.\ 2018, \mnras, 479, 2, 2464
\bibitem[Gierli{\'n}ski \& Done(2004)]{2004MNRAS.349L...7G} Gierli{\'n}ski, M. \& Done, C.\ 2004, \mnras, 349, 1, L7
\bibitem[Gliozzi \& Williams(2020)]{2020MNRAS.491..532G} Gliozzi, M. \& Williams, J.~K.\ 2020, \mnras, 491, 1, 532
\bibitem[Goodrich(1989)]{1989ApJ...342..224G} Goodrich, R.~W.\ 1989, \apj, 342, 224
\bibitem[Graham et al.(2020)]{2020MNRAS.491.4925G} Graham, M.~J., Ross, N.~P., Stern, D., et al.\ 2020, \mnras, 491, 4, 4925.
\bibitem[Green et al.(2022)]{2022ApJ...933..180G} Green, P.~J., Pulgarin-Duque, L., Anderson, S.~F., et al.\ 2022, \apj, 933, 2, 180. 
\bibitem[Grupe et al.(2010)]{2010ApJS..187...64G} Grupe, D., Komossa, S., Leighly, K.~M., et al.\ 2010, \apjs, 187, 1, 64
\bibitem[Gu \& Chen(2010)]{2010AJ....139.2612G} Gu, M. \& Chen, Y.\ 2010, \aj, 139, 6, 2612
\bibitem[Gu et al.(2015)]{2015ApJS..221....3G} Gu, M., Chen, Y., Komossa, S., et al.\ 2015, \apjs, 221, 1, 3
\bibitem[Guo et al.(2018)]{2018ascl.soft09008G} Guo, H., Shen, Y., \& Wang, S.\ 2018, Astrophysics Source Code Library. ascl:1809.008
\bibitem[Gupta et al.(2024)]{2024A&A...691A.203G} Gupta, K.~K., Ricci, C., Temple, M.~J., et al.\ 2024, \aap, 691, A203
\bibitem[Helfand et al.(2015)]{2015ApJ...801...26H} Helfand, D.~J., White, R.~L., \& Becker, R.~H.\ 2015, \apj, 801, 1, 26
\bibitem[HI4PI Collaboration et al.(2016)]{2016A&A...594A.116H} HI4PI Collaboration, Ben Bekhti, N., Flöer, L., et al.\ 2016, \aap, 594, A116
\bibitem[Ho(2008)]{2008ARA&A..46..475H} Ho, L.~C.\ 2008, \araa, 46, 475
\bibitem[Husemann et al.(2016)]{2016A&A...593L...9H} Husemann, B., Urrutia, T., Tremblay, G.~R., et al.\ 2016, \aap, 593, L9
\bibitem[Jiang et al.(2025)]{2025MNRAS.539..832J} Jiang, J., Walton, D.~J., Gallo, L.~C., et al.\ 2025, \mnras, 539, 2, 832
\bibitem[Just et al.(2007)]{2007ApJ...665.1004J} Just, D.~W., Brandt, W.~N., Shemmer, O., et al.\ 2007, \apj, 665, 2, 1004
\bibitem[Kellermann et al.(1989)]{1989AJ.....98.1195K} Kellermann, K.~I., Sramek, R., Schmidt, M., et al.\ 1989, \aj, 98, 1195
\bibitem[Khachikian \& Weedman(1974)]{1974ApJ...192..581K} Khachikian, E.~Y. \& Weedman, D.~W.\ 1974, \apj, 192, 581
\bibitem[Komossa(2008)]{2008RMxAC..32...86K} Komossa, S.\ 2008, Revista Mexicana de Astronomia y Astrofisica Conference Series, 32, 86
\bibitem[Komossa et al.(2015)]{2015A&A...574A.121K} Komossa, S., Xu, D., Fuhrmann, L., et al.\ 2015, \aap, 574, A121
\bibitem[Komossa(2018)]{2018rnls.confE..15K} Komossa, S.\ 2018, Revisiting Narrow-Line Seyfert 1 Galaxies and their Place in the Universe, 15
\bibitem[Komossa et al.(2006)]{2006AJ....132..531K} Komossa, S., Voges, W., Xu, D., et al.\ 2006, \aj, 132, 2, 531
\bibitem[Kynoch et al.(2023)]{2023MNRAS.520.2781K} Kynoch, D., Mitchell, J.~A.~J., Ward, M.~J., et al.\ 2023, \mnras, 520, 2, 2781
\bibitem[Laki{\'c}evi{\'c} et al.(2018)]{2018MNRAS.478.4068L} Laki{\'c}evi{\'c}, M., Popovi{\'c}, L. {\v{C}}., \& Kova{\v{c}}evi{\'c}-Doj{\v{c}}inovi{\'c}, J.\ 2018, \mnras, 478, 3, 4068
\bibitem[Leighly(1999)]{1999ApJS..125..317L} Leighly, K.~M.\ 1999, \apjs, 125, 2, 317
\bibitem[Li \& Gu(2018)]{2018MNRAS.481L..45L} Li, S.-L. \& Gu, M.\ 2018, \mnras, 481, 1, L45
\bibitem[Li(2019)]{2019MNRAS.490.3793L} Li, S.-L.\ 2019, \mnras, 490, 3, 3793
\bibitem[Li et al.(2008)]{2008ApJ...688..826L} Li, Z.-Y., Wu, X.-B., \& Wang, R.\ 2008, \apj, 688, 2, 826
\bibitem[Liao et al.(2020)]{2020MNRAS.497..482L} Liao, M., Gu, M., Zhou, M., et al.\ 2020, \mnras, 497, 1, 482
\bibitem[Liao et al.(2024)]{2024MNRAS.528.3696L} Liao, M., Wang, J., Ren, W., et al.\ 2024, \mnras, 528, 2, 3696
\bibitem[Lu \& Wang(2025)]{2025ApJ...994..154L} Lu, W.-T. \& Wang, J.-X.\ 2025, \apj, 994, 2, 154.
\bibitem[Lusso et al.(2010)]{2010A&A...512A..34L} Lusso, E., Comastri, A., Vignali, C., et al.\ 2010, \aap, 512, A34
\bibitem[Lyu et al.(2021)]{2021MNRAS.506.4188L} Lyu, B., Yan, Z., Yu, W., et al.\ 2021, \mnras, 506, 3, 4188
\bibitem[Massaro et al.(2009)]{2009A&A...495..691M} Massaro, E., Giommi, P., Leto, C., et al.\ 2009, \aap, 495, 2, 691
\bibitem[Massaro et al.(2015)]{2015Ap&SS.357...75M} Massaro, E., Maselli, A., Leto, C., et al.\ 2015, \apss, 357, 1, 75
\bibitem[Merloni et al.(2003)]{2003MNRAS.345.1057M} Merloni, A., Heinz, S., \& di Matteo, T.\ 2003, \mnras, 345, 4, 1057
\bibitem[Miller et al.(2011)]{2011ApJ...726...20M} Miller, B.~P., Brandt, W.~N., Schneider, D.~P., et al.\ 2011, \apj, 726, 1, 20
\bibitem[Mundo et al.(2020)]{2020MNRAS.496.2922M} Mundo, S.~A., Kara, E., Cackett, E.~M., et al.\ 2020, \mnras, 496, 3, 2922
\bibitem[Ojha et al.(2020)]{2020ApJ...896...95O} Ojha, V., Chand, H., Dewangan, G.~C., et al.\ 2020, \apj, 896, 2, 95
\bibitem[Osterbrock \& Pogge(1985)]{1985ApJ...297..166O} Osterbrock, D.~E. \& Pogge, R.~W.\ 1985, \apj, 297, 166
\bibitem[Paliya et al.(2018)]{2018ApJ...853L...2P} Paliya, V.~S., Ajello, M., Rakshit, S., et al.\ 2018, \apjl, 853, 1, L2
\bibitem[Paliya et al.(2024)]{2024MNRAS.527.7055P} Paliya, V.~S., Stalin, C.~S., Dom{\'\i}nguez, A., et al.\ 2024, \mnras, 527, 3, 7055
\bibitem[Paliya et al.(2019)]{2019ApJ...872..169P} Paliya, V.~S., Parker, M.~L., Jiang, J., et al.\ 2019, \apj, 872, 2, 169
\bibitem[Panessa et al.(2019)]{2019NatAs...3..387P} Panessa, F., Baldi, R.~D., Laor, A., et al.\ 2019, Nature Astronomy, 3, 387
\bibitem[Panessa et al.(2011)]{2011MNRAS.417.2426P} Panessa, F., de Rosa, A., Bassani, L., et al.\ 2011, \mnras, 417, 4, 2426
\bibitem[Potts \& Villforth(2021)]{2021A&A...650A..33P} Potts, B. \& Villforth, C.\ 2021, \aap, 650, A33. 
\bibitem[Rakshit et al.(2017)]{2017ApJS..229...39R} Rakshit, S., Stalin, C.~S., Chand, H., et al.\ 2017, \apjs, 229, 2, 39
\bibitem[Ren et al.(2024)]{2024ApJ...974..153R} Ren, W., Guo, H., Shen, Y., et al.\ 2024, \apj, 974, 2, 153
\bibitem[Rosa et al.(2025)]{2025A&A...698A.160R} Rosa, V., Foschini, L., \& Ciroi, S.\ 2025, \aap, 698, A160
\bibitem[Schlegel et al.(1998)]{1998ApJ...500..525S} Schlegel, D.~J., Finkbeiner, D.~P., \& Davis, M.\ 1998, \apj, 500, 2, 525
\bibitem[Shang et al.(2011)]{2011ApJS..196....2S} Shang, Z., Brotherton, M.~S., Wills, B.~J., et al.\ 2011, \apjs, 196, 1, 2
\bibitem[Shemmer et al.(2008)]{2008ApJ...682...81S} Shemmer, O., Brandt, W.~N., Netzer, H., et al.\ 2008, \apj, 682, 1, 81
\bibitem[Shen et al.(2011)]{2011ApJS..194...45S} Shen, Y., Richards, G.~T., Strauss, M.~A., et al.\ 2011, \apjs, 194, 2, 45
\bibitem[Tanaka et al.(2004)]{2004PASJ...56L...9T} Tanaka, Y., Boller, T., Gallo, L., et al.\ 2004, \pasj, 56, L9
\bibitem[Urry \& Padovani(1995)]{1995PASP..107..803U} Urry, C.~M. \& Padovani, P.\ 1995, \pasp, 107, 803
\bibitem[V{\'e}ron-Cetty \& V{\'e}ron(2006)]{2006A&A...455..773V} V{\'e}ron-Cetty, M.-P. \& V{\'e}ron, P.\ 2006, \aap, 455, 2, 773
\bibitem[Vestergaard \& Wilkes(2001)]{2001ApJS..134....1V} Vestergaard, M. \& Wilkes, B.~J.\ 2001, \apjs, 134, 1, 1
\bibitem[Viswanath et al.(2019)]{2019ApJ...881L..24V} Viswanath, G., Stalin, C.~S., Rakshit, S., et al.\ 2019, \apjl, 881, 1, L24
\bibitem[Waddell \& Gallo(2020)]{2020MNRAS.498.5207W} Waddell, S.~G.~H. \& Gallo, L.~C.\ 2020, \mnras, 498, 4, 5207
\bibitem[Wang et al.(2025)]{2025ApJ...987L..26W} Wang, A., An, T., Kellermann, K.~I., et al.\ 2025, \apjl, 987, 2, L26
\bibitem[Wang \& Netzer(2003)]{2003A&A...398..927W} Wang, J.-M. \& Netzer, H.\ 2003, \aap, 398, 927
\bibitem[Wang et al.(2006)]{2006ApJ...645..890W} Wang, R., Wu, X.-B., \& Kong, M.-Z.\ 2006, \apj, 645, 2, 890
\bibitem[Webb et al.(2020)]{2020A&A...641A.136W} Webb, N.~A., Coriat, M., Traulsen, I., et al.\ 2020, \aap, 641, A136
\bibitem[White et al.(1997)]{1997ApJ...475..479W} White, R.~L., Becker, R.~H., Helfand, D.~J., et al.\ 1997, \apj, 475, 2, 479
\bibitem[Xu et al.(2012)]{2012AJ....143...83X} Xu, D., Komossa, S., Zhou, H., et al.\ 2012, \aj, 143, 4, 83
\bibitem[Xu et al.(2021)]{2021MNRAS.507.3572X} Xu, X., Ding, N., Gu, Q., et al.\ 2021, \mnras, 507, 3, 3572
\bibitem[Yao \& Komossa(2023)]{2023MNRAS.523..441Y} Yao, S. \& Komossa, S.\ 2023, \mnras, 523, 1, 441
\bibitem[Yip et al.(2004a)]{2004AJ....128..585Y} Yip, C.~W., Connolly, A.~J., Szalay, A.~S., et al.\ 2004, \aj, 128, 2, 585
\bibitem[Yip et al.(2004b)]{2004AJ....128.2603Y} Yip, C.~W., Connolly, A.~J., Vanden Berk, D.~E., et al.\ 2004, \aj, 128, 6, 2603
\bibitem[Yuan et al.(2008)]{2008ApJ...685..801Y} Yuan, W., Zhou, H.~Y., Komossa, S., et al.\ 2008, \apj, 685, 2, 801
\bibitem[Zhou \& Gu(2021)]{2021RAA....21....4Z} Zhou, M.-H. \& Gu, M.-F.\ 2021, Research in Astronomy and Astrophysics, 21, 1, 004
\bibitem[Zhou et al.(2026)]{2026MNRAS.548ag588Z} Zhou, M., Wu, X., Xu, L., et al.\ 2026, \mnras, 548, 3, stag588
\bibitem[Zhou \& Gu(2020)]{2020ApJ...893...39Z} Zhou, M. \& Gu, M.\ 2020, \apj, 893, 1, 39
\bibitem[Zhou et al.(2023)]{2023MNRAS.519..909Z} Zhou, M., Gu, M., Liao, M., et al.\ 2023, \mnras, 519, 1, 909
\bibitem[Zhou \& Zhang(2010)]{2010ApJ...713L..11Z} Zhou, X.-L. \& Zhang, S.-N.\ 2010, \apjl, 713, 1, L11
\bibitem[Zhu et al.(2020)]{2020MNRAS.496..245Z} Zhu, S.~F., Brandt, W.~N., Luo, B., et al.\ 2020, \mnras, 496, 1, 245

\end{thebibliography}
\end{document}